\documentclass[aps, prd, nofootinbib]{revtex4}

\newcommand{\bb}{\textbf}
\newcommand{\bbb}{\bigskip}
\newcommand{\be}{\begin{eqnarray}}
\newcommand{\bee}{\begin{enumerate}}
\newcommand{\bit}{\begin{itemize}}

\def\bkt#1{\left(#1\right)} 
\def\bkts#1{\left[#1\right]} 
\def\bkta#1{\langle#1\rangle} 

\def\ee{\end{eqnarray}}
\newcommand{\eee}{\end{enumerate}}
\def\eg{\textit{e.g.} }
\newcommand{\eit}{\end{itemize}}
\def\etal{\textit{et al.}} 

\def\fnl{f_{\mbox{\scriptsize NL}}}
\def\ff{\phantom{.}}
\def\fsky{f_{\mbox{\scriptsize sky}}}

\def\gnl{g_{\mbox{\scriptsize NL}}}

\def\iee{\textit{i.e. }}
\newcommand{\ii}{\textit}

\def\lab{\label}

\newcommand{\mb}{\mathbf}
\def\mc#1{\mathcal{#1}}
\newcommand{\mmm}{\medskip}

\newcommand{\nn}{\nonumber}

\def\pr{\prime}
\def\re#1{(\ref{#1})}

\newcommand{\sss}{\smallskip}
\def\sub#1{_{\mbox{\scriptsize{#1}}}}
\def\sun{\odot}
\def\super#1{^{\mbox{\scriptsize{#1}}}}

\newcommand\apjl{ApJ}%
\newcommand\apjs{ApJS}%

\newcommand\aap{A\&A}%
\usepackage{amsmath}
\usepackage{amsfonts}
\usepackage{amssymb}
\usepackage{amsthm}
\usepackage{graphicx}
\usepackage{epsfig}
\usepackage{hyperref}
\usepackage[justification=raggedright,font=small]{caption}
\usepackage{subfig}

\usepackage{bm}

\begin{document}

\title{Primordial Non-Gaussianity and \\Extreme-Value Statistics of Galaxy Clusters}

\author{Sirichai Chongchitnan}
\email{siri@astro.ox.ac.uk}

\author{Joseph Silk}
\affiliation{Department of Physics, University of Oxford and \\
Beecroft Institute for Particle Astrophysics and Cosmology,  \\ Denys Wilkinson Building, 1 Keble Road, Oxford, OX1 3RH, United Kingdom.}

\begin{abstract}

What is the size of the \ii{most} massive object one expects to find in a survey of a given volume? In this paper, we present a solution to this problem using Extreme-Value Statistics, taking into account primordial non-Gaussianity and its effects on the abundance and the clustering of rare objects. We calculate the probability density function (pdf) of extreme-mass clusters in a survey volume, and show how primordial non-Gaussianity shifts the peak of this pdf. We also study the sensitivity of the extreme-value pdfs to changes in the mass functions, survey volume, redshift coverage and the normalization of the matter power spectrum, $\sigma_8$. For `local' non-Gaussianity parametrized by $\fnl$, our correction for the extreme-value pdf due to the bias is important when $\fnl\gtrsim100$, and becomes more significant for wider and deeper surveys. Applying our formalism to the massive high-redshift cluster XMMUJ0044.0-2-33, we find that its existence is consistent with $\fnl=0$, although the conclusion is sensitive to the assumed values of $\fsky$ and $\sigma_8$. We also discuss the convergence of the extreme-value distribution to one of the three possible asymptotic forms, and argue that the convergence is insensitive to the presence of non-Gaussianity. 

\end{abstract}


\maketitle

\section{Introduction}

The statistics of the primordial seeds that grew into the observed large-scale structures holds a wealth of information about the physics of the primordial Universe. In the simplest models of inflation, the primordial density fluctuations obey an almost Gaussian statistics (see \cite{bartolo} for a review). Tiny deviations from Gaussianity may be quantified, amongst other ways\footnote{In general, a non-Gaussian pdf can have divergent moments (e.g. the Cauchy distribution). In this work we assume that the primordial density distribution has finite moments up to third order (\ii{i.e.} finite skewness).}, by the `local' non-Gaussianity parameter, $\fnl$, defined via the expansion of the non-linear Newtonian potential
\be \Phi = \phi + \fnl(\phi^2-\bkta{\phi^2})+\ldots,\ee
where $\phi$ is a Gaussian random field. This form of non-Gaussianity arises in simple models of single and multi-field inflation \citep{maldacena,rigopoulos,byrnes} as well as some curvaton models \citep{bartolo2,sasaki,malik}. Observational constraints on $\fnl$  from the cosmic microwave background (CMB) anisotropies are currently consistent with $\fnl=32\pm42$ ($2\sigma$) \citep{komatsu}. However, if $\fnl$ is in fact much smaller, its effects on the CMB would be difficult to extract and distinguished from non-Gaussianity arising from secondary sources such as gravitational lensing and instrumental noise \cite{cooray,yu}.



The statistics of large-scale structures offers a complementary probe of non-Gaussianity on much smaller scales than the CMB. Indeed, it is possible that $\fnl$ measured on Gpc scales may be quite different from that measured on Mpc scales. In wavenumber space, this translates to a possible $k$-dependence of $\fnl$, which have been hinted at by the numerous observations of massive high-redshift clusters \cite{jee,stott,brodwin,santos,tanaka}. These massive clusters exist, according to some, in greater abundances than expected from a Gaussian statistics. Some authors have concluded that the level of non-Gaussianity on Mpc scale required to explain the existence of certain rare clusters is $\fnl=$ a few $\times10^2$ \cite{enqvist,hoyle}. In contrast, some have argued that these claims are based on misinterpretation of data, and that the occurrences of these rare objects are in fact consistent with a Gaussian statistics \cite{hotchkiss, cayon,mortonson}. 

In this work, we offer our opinion on this debate by presenting an approach to calculating the probability of observing rare objects based on extreme-value statistics. We begin by asking: what is the probability distribution of the \ii{most} massive clusters found within a given volume at a given redshift range? Our technique relies on a basic application of the so-called void probability distribution introduced by White \cite{white}. This approach was successfully used to study the abundances of massive clusters given a Gaussian initial condition in \cite{davis,colombi}. In this work, we extend the groundwork laid by these authors to study the effect of $\fnl$ on the distribution of extreme-mass objects. For other cosmological applications of extreme-value theory, see \cite{bernstein,antal,dobos,bhavsar, mikelsons,sheth,waizmann,waizmann2}.

Previous approaches to extreme-value statistics of clusters have so far either neglected the clustering, or bias, of galaxy clusters \cite{waizmann2, harrison}, or considered it in the Gaussian case \cite{davis,sheth}. In this work, however, we have included the effects of the bias in the presence of non-Gaussianity. Whilst Davis \ii{et al.} \cite{davis} have previously reported that the effects of the bias on the extreme-value distribution are small in the Gaussian case, it remains to be shown if this also holds in the presence of non-Gaussianity, which can introduce a strong scale dependence in the bias \cite{dalal}. We investigate this problem in this work using the formalism of Valageas \cite{valageasA,valageasB}, who showed how the bias can be calculated in real space for a given $\fnl$. As we shall see later, the contribution from the bias can indeed be significant if $\fnl$ is sufficiently large.

\section{The primordial density fluctuations}

In this section, we introduce the parameters needed to describe the primordial density fluctuations statistically. Some of our present conventions (such as the Fourier transform and the moments of the density fluctuations) slightly differ from our earlier work \cite{me8,me9}. In particular, smoothing by a window function will be kept explicit, in contrast with other work in which overdensities are defined to be implicitly smoothed. 

Let $\rho_c$, $\rho_b$, $\rho_r$, $\rho_\Lambda$ be the time-dependent energy densities of cold dark matter, baryons, radiation and dark energy. Let $\rho_m=\rho_c+\rho_b$. We define the density parameter for species $i$ as 
\be \Omega_i \equiv {\rho_i (z=0)\over \rho\sub{crit}},\ee
where $\rho\sub{crit}$ is the critical density defined as $\rho\sub{crit}\equiv 3H_0^2/8\pi G $. The Hubble constant, $H_0$, is parametrized by $h$ via the usual formula $H_0\equiv100h \mbox{ km\ff s}^{-1}\mbox{Mpc}^{-1}$. Results from a range of astrophysical observations are consistent with $h\simeq 0.7$, $\Omega_c\simeq0.23$, $\Omega_b\simeq0.046$, $\Omega_r\simeq 8.6\times10^{-5}$ and $\Omega_\Lambda=1-\Omega_m-\Omega_r$ (see \eg \cite{komatsu,lahav+}). 

The density fluctuation field, $\delta$, is defined at redshift $z$ as
\be \delta(\mb{x},z)\equiv {\rho_m(\mb{x},z)-\bkta{\rho_m(z)}\over\bkta{\rho_m(z)}},\ee
where $\bkta{\rho_m}$ is the mean matter energy density. To make the notation less cumbersome, we shall write $\delta(\mb{x})$ to mean $\delta(\mb{x},z)$. The Fourier decomposition of $\delta(\mb{x})$ is given by 
\be  \delta(\mb{x})=\int d{\mb k} \ff \delta(\mb{k}) e^{i\mb{k}\cdot\mb{x}}.\ee 

The gravitational Newtonian potential, $\Phi$, is related to the density fluctuation by the cosmological Poisson equation
\be \delta(\mb{k}) &=&  \mc{A}(k,z)\Phi(\mb{k}),\\
\mc{A}(k,z)&\equiv&{2\over3\Omega_m }\bkt{k\over H_0}^2 T(k\sub{EH}) D(z),\ee where $T$ is the transfer function and $D$ is the linear growth factor calculated using the fitting formula of \cite{carroll,lahav} with $D(0)\approx0.76$. We follow the approach outlined in \cite{weinberg} and use the transfer function 

\begin{align} 
T(k)={\ln[1+(0.124k)^2]\over (0.124k)^2}\bkts{1+(1.257k)^2+(0.4452k)^4+(0.2197k)^6 \over1+(1.606k)^2+(0.8568k)^4+(0.3927k)^6  }^{1/2}.
\end{align}
In addition, we also incorporate the baryonic correction of Eisenstein and Hu \cite{eisenstein}, whereby the transfer function is evaluated at 
\be k\sub{EH}={k\Omega_r^{1/2}\over H_0\Omega_m}\bkts{\alpha+{{1-\alpha}\over{1+(0.43ks)^4}}}^{-1},\lab{EH}\ee
with 
$$\alpha=1-0.328\ln(431\Omega_mh^2){\Omega_b\over\Omega_m}+0.38\ln(22.3\Omega_mh^2)\bkt{\Omega_b\over\Omega_m}^2,$$
and 
$$s={44.5\ln(9.83/\Omega_m h^2) \over \sqrt{1+10(\Omega_b h^2)^{3/4}}} \ff \mbox{Mpc}.$$


The power spectrum, $P(k)$, can be defined via the two-point correlation function in Fourier space as
\be \bkta{\delta(\mb{k_1}),\delta(\mb{k_2})}=\delta_D(\mb{k_1}+\mb{k_2})P(k),\ee
where $\delta_D$ is the 3-dimensional Dirac delta function. In linear perturbation theory, it is usually assumed that inflation laid down a primordial  spectrum of the form $P(k)\propto k^{n_s}$, where $n_s$ is the scalar spectral index (assumed to be 0.96 in this work). 

The variance of linear density fluctuations smoothed on scale $R$ is given by
\be \sigma^2_R= 4\pi\int_0^\infty {dk\over k}  W^2(kR)\mc{P}(k)\lab{vari}.\ee  
where $\mc{P}(k)\equiv {k^3} P(k)\propto [\mc{A}(k,z)]^2\bkt{k/ H_0}^{n_s-1}.$ 

We choose $W$ to be the spherical top-hat function of radius $R$. In Fourier space, we have 
\be W(kR)=3\bkts{{\sin(kR)\over (kR)^3}-{\cos(kR)\over (kR)^2}}.\ee
The mass, $M$, of matter enclosed by a top-hat window of radius $R$ is given by
\be M\equiv {4\over3}\pi R^3\rho_m\approx 1.16\times10^{12}\Omega_m\bkt{{R\over h^{-1}\mbox{Mpc}}}^3 \ff h^{-1}M_\sun.\ee
With the above relation, the smoothed variance, $\sigma_R$, can be equivalently expressed as $\sigma_M$. Finally, the normalization of $\mc{P}(k)$ is such that 
\be \sigma_8\equiv\sigma(R=8h^{-1}\mbox{Mpc},z=0)=0.801.\lab{sig8}\ee



\section{Cluster number counts}\lab{massfun}

The mean number density, $n$, of objects with mass greater than $m$, at redshift $z$ can be calculated by
\be n(>m,z)= \int_{m}^{\infty} {dn\over dM}\, dM,\lab{en}\ee
where $dn/dM$ is the differential number density. In the presence of local non-Gaussianity, Matarrese, Verde and Jimenez \cite{mvj} used a saddle-point expansion (assuming that the deviation from Gaussianity is sufficiently weak for such an expansion to converge) to derive a correction factor for $dn/dM$ of the form 
\be \mc{R}&=& \exp\bkt{S_3 \delta_c^3\over 6\sigma_M^2}\bkts{{\delta_c^2\over 6\Delta}\cdot{dS_3\over d\ln \sigma_M} + \Delta},\\
\Delta&\equiv& \sqrt{1-{\delta_c S_3\over3}},\lab{mvj}\ee
where the third cumulant, $S_3$, is given by $S_3=\bkta{\delta^3}$, and is assumed  to be almost independent of the smoothing mass-scale, $M$. This latter assumption indeed holds on cluster scales (see \re{skewme} below). $S_3$ can be calculated either from a 3-dimensional integral
\be S_3(M) ={6f_{R;RR}(0)\over \sigma^4_M},\lab{skewv}\ee with $f_{R;RR}$ defined in equation \re{fff}, or from the fitting formula \cite{me8}
\be S_3(M) = {3.15\times 10^{-4}\fnl\over \sigma_M^{0.838}}.\lab{skewme}\ee 
The critical overdensity, $\delta_c$, is taken to be\footnote{Note that $\delta_c$ is taken to be constant, whilst the redshift dependence is carried by the factor $\mc{A}(k,z)$. This convention agrees with \cite{valageasA,valageasB} but is different from the ``excursion-set" convention in which the redshift dependence is carried by $\delta_c$, with $\sigma$ extrapolated to $z=0$ (e.g. \cite{grossi}). We believe our present convention will facilitate comparison with \cite{valageasA,valageasB}, whose results will be used in the next section.}
\be \delta_c=\sqrt{a}\times 1.686, \ee with the `fudge factor' $\sqrt{a}=0.9$ as recommended by \cite{wagner}, although there is still debate over its value \cite{pillepich,grossi}.

There are alternative forms of the correction factor, $\mc{R}$, given by LoVerde \etal \ff \cite{loverde} based on a low-order Edgeworth expansion, and by Paranjape \etal \ff \cite{paranjape} based on resumming terms in the saddle-point expansion of the mass function. We tested both of these alternative corrections and found that, in the range of parameters used in this paper, there are only small differences between the various prescriptions and our main results are unaffected by the choice of the correction factor. In the rest of this work, we shall use only the MVJ correction factor (see \cite{wagner} for a comparison between the correction factors).


In summary, we shall consider the non-Gaussian differential abundance of the form \be {dn\over dM} = \mc{R}\times F(\nu){\rho_m\over M} {d \ln\sigma^{-1}\over dM},\lab{dndm}\ee
where $\nu\equiv \delta_c/\sigma_M$ and $F(\nu)$ is one of the following three standard mass-functions
\be
\mbox{Press-Schechter \cite{ps}}&& \qquad F\sub{PS} = \sqrt{2\over\pi}\nu e^{-\nu^2/2},\\
\mbox{Sheth-Tormen \cite{st}}&& \qquad F\sub{ST} = 0.322\sqrt{2a\over\pi}\nu\exp\bkt{-{a\nu^2\over2}}\bkts{1+\bkt{a\nu^2}^{-0.3}},\quad a= 0.707,\\
\mbox{Tinker \etal\ff \cite{tinker,tinker2}}&& \qquad F\sub{Tinker} = 0.368\bkts{1+\bkt{\beta\nu}^{-2\phi}}\nu^{2\eta+1}e^{-\gamma\nu^2/2},\\
&& \qquad \beta = 0.589(1+z)^{0.2}, \phi=-0.729(1+z)^{-0.08},\nn\\
&& \qquad \eta = -0.243(1+z)^{0.27}, \gamma = 0.864(1+z)^{-0.01}.\nn
\ee
The Press-Schechter and Sheth-Tormen mass functions can be derived by considering the overdensity, $\delta$, as a stochastic function of the smoothing mass scale, $M$, and associating trajectories (in the $(M,\delta)$ plane) that overshoot a barrier, $\delta_c$, with a collapsed object. One can show that a spherical collapse can be associated with a barrier of constant height, resulting in the Press-Schechter mass function \cite{ps}, whilst an ellipsoidal collapse can be  associated with a drifting barrier, $\delta_c=1.686(1+\alpha(\delta/\sigma)^\beta)$ ($\alpha$,$\beta$ constant), giving the Sheth-Tormen mass function \cite{st}.  The Tinker mass function belongs to a family of so-called universal mass functions derived from a suite of $N$-body simulations, with the functional form deviating from simulation results by $\lesssim5\%$ in the redshift range considered here (for detail see \cite{tinker,tinker2}, and also \cite{jenkins,warren,reed}).

\begin{figure}
\centering
\includegraphics[width= 6.2cm, angle = -90]{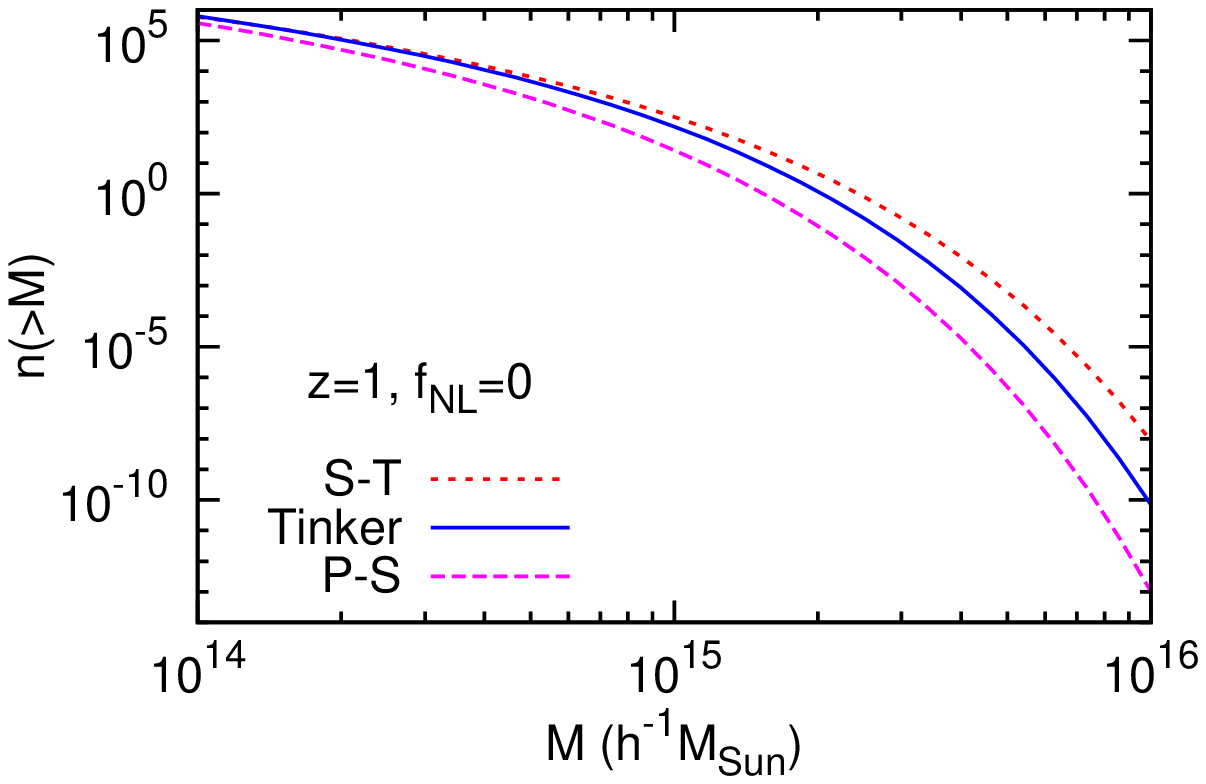}
\includegraphics[width= 6.2cm, angle = -90]{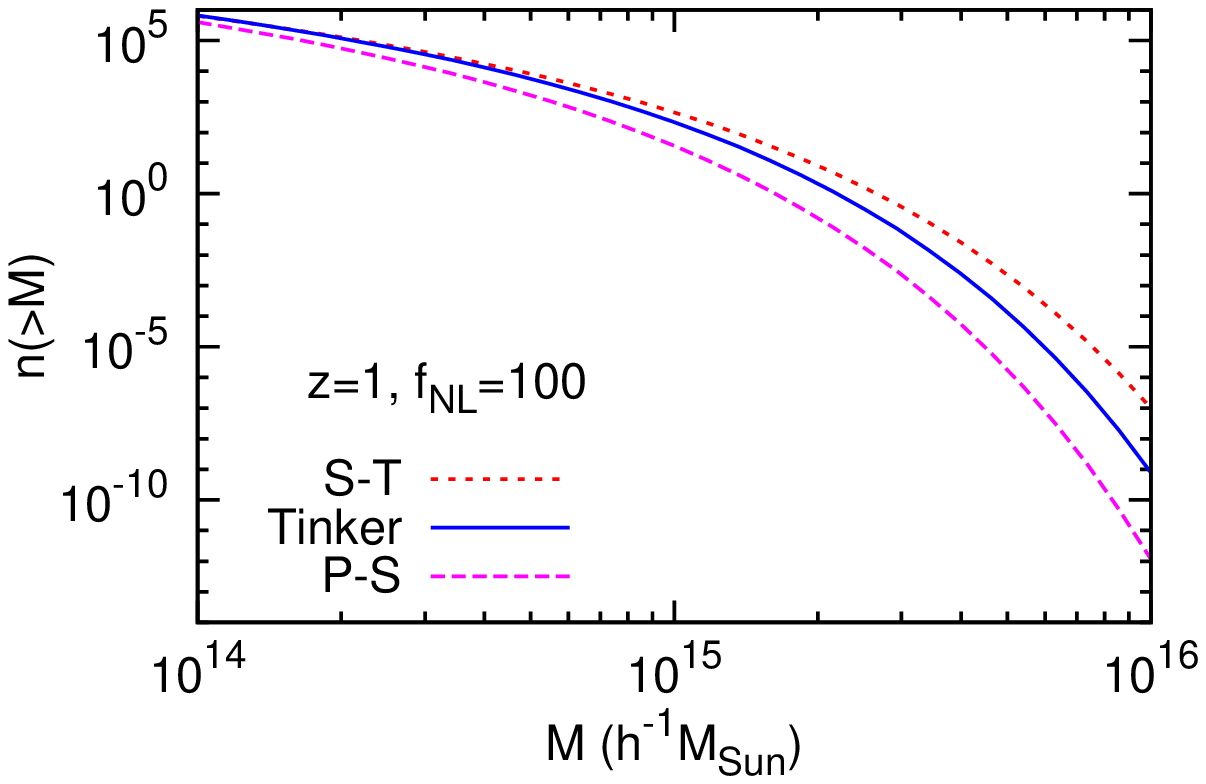}
\caption{The number of objects above mass $M$ with $\fnl=0$ (left) and 100 (right) at $z=1$, calculated over the full sky using the Press-Schechter (P-S), Sheth-Tormen (S-T) and Tinker \etal \ff mass functions. With $\fnl=100$, the number count increases by roughly an order of magnitude at the high-mass end compared with $\fnl=0$.}
\label{fign}

\end{figure}

The number of objects with mass above $M$ expected at redshift $z$ is given by the integral
\be {dV\over dz}\times\int_M^\infty {dn\over dm}dm,\ee 
where the volume element $dV/dz$ satisfies
\be {dV\over dz}&=&f\sub{sky} {4\pi\over H(z)}\bkt{\int_0^{z} {dz^\pr\over H(z^\pr)}}^2,\lab{surveyvol}\\
H(z)&\approx&H_0\bkts{\Omega_m(1+z)^3+\Omega_\Lambda}^{1/2},
\ee
and $\fsky$ is the fraction of the sky covered by the survey. The number count for $z=1$, $\fsky=1$ and $\fnl=0$ or $100$ is shown in figure \ref{fign}. Comparing the mass functions, we see that the Sheth-Tormen gives the highest number count, followed by the Tinker and the Press-Schechter mass functions. Changing $\fnl$ to $100$ (right panel) increases the number count at the high-mass end by roughly an order of magnitude. See \eg \cite{pillepich,bhattacharya} for more comparisons between various mass functions.


\section{Bias}\lab{bias}

In the seminal work of Dalal \etal \cite{dalal}, it was shown quantitatively how  non-Gaussianity gives rise to characteristic changes in the clustering of density peaks corresponding to rare objects. At leading order, it is common to define the bias in Fourier space as the ratio of the power-spectra
\be b^2(k) = {P\sub{halo}(k)\over P_m(k)},\lab{biask}\ee
which represents the amplitude at which density peaks ($P\sub{halo}$) trace the underlying dark matter distribution ($P_{m}$). The Fourier space formalism was used by a majority of papers on non-Gaussian bias (\eg \cite{matarrese,wagner,desjacques,schmidt,shandera}).

However, an arguably more intuitive measure of the bias is in real space, where the density fluctuation in peaks (\iee luminous objects) is expressed as a non-linear function of the local dark-matter density fluctuation. 
On linear scales, the bias is given by the ratio of the correlation functions \cite{fry,kaiser}
\be b^2(r) = {\xi\sub{pk}(r)\over\xi(r)},\lab{biasr}\ee
where $r$ is the comoving length in Eulerian space (throughout this work  quantities with a subscript `pk' are associated with density peaks). The correlation function, $\xi$, is defined as
\be \xi(\mb{x_1},\mb{x_2})=\bkta{\delta(\mb{x_1}),\delta(\mb{x_2})},\quad r= |\mb{x_1}-\mb{x_2}|.\ee On linear scales where $\mc{P}(k)$ is of a power-law form parametrized by $n\sub{s}$, we can write
\be  \xi(r)=4\pi\int_0^\infty {dk\over k} \ff  \mc{P}(k) j_0(kr), \lab{xi}\ee
where $j_0(x)=\sin x/x$. The real-space bias tells us directly about the clustering amplitude of density peaks separated by distance $r$. We shall refer to $r$ as the separation length.

Unfortunately, when comparing \re{biask} and \re{biasr}, we see that the real-space bias, $b(r)$, and the Fourier-space bias, $b(k)$, are not simply related via a Fourier transform but rather a complicated convolution. In \cite{me9}, we avoided  this problem by interpreting \re{biasr} as a ratio of joint probabilities of finding overdensities at two points distance $r$ apart, and then applying a bivariate Edgeworth expansion. Due to the algebraic nature of the Edgeworth expansion, this technique was readily applied to non-Gaussianity parametrized by the cubic order parameter, $\gnl$, but surprisingly the application is much less straightforward for $\fnl$.

An alternative method for calculating the real-space bias in the presence of $\fnl$ was presented by Valageas \cite{valageasA,valageasB} in which he showed that analytic calculations could be made as long as the separation length is sufficiently large. In this work, we shall follow this formalism, of which we give a simplified account here. 

A crucial element in the real-space approach is the mapping between the separation length, $s$, in Lagrangian coordinates (associated with linear density fluctuations) and that in Eulerian coordinates (associated with non-linear fluctuations). This relation is given by
\be s \simeq r\bkt{1+{2\delta_R(r)\over3}},\ee
accurate at large distances where $\delta_R(r)\ll1$. Here $\delta_R(r)$ can be interpreted as the radial profile of the linear density contrast from the centre of the halo. The profile is given by
\be \delta_R(r)={\delta_c\over \sigma^2_R}\ff\sigma^2_{R,0}(r)+{\delta_c^2\over \sigma_R^4}\ff\bkts{f_{0;RR}(r)+2g_{R;0R}(r)-3{\sigma^2_{R,0}(r)\over \sigma_R^2}f_{R;RR}(0)}.\ee
In this expression, the functions $\sigma_{R_1,R_2}(r)$, $f_{R;R_1R_2}(r)$ and $g_{R;R_1R_2}(r)$ are defined by the following integrals\footnote{In this paper $\fnl$ is defined in the `LSS' convention. The `CMB' convention, as used in \cite{valageasA}, satisfies $\fnl\super{CMB}=D(0)\fnl\super{LSS}$.}

\be {\sigma}^2_{R_1,R_2}(r)&=&4\pi\int_0^\infty {dk\over k}\mc{P}(k)W(kR_1)W(kR_2)j_0(kr),\\
f_{R;R_1R_2}(r)&=& 8\pi^2D(0)\fnl\int_0^\infty {dk_1\over k_1} {\mc{P}(k_1)}W(k_1R_1) \int_0^\infty {dk_2\over k_2} {\mc{P}(k_2)}W(k_2R_2)\int_{-1}^{1}d\mu W(kR) {\mc{A}(k)\over \mc{A}(k_1)\mc{A}(k_2)} j_0(kr), \lab{fff} \\
g_{R;R_1R_2}(r)&=& 8\pi^2D(0)\fnl\int_0^\infty  {dk_1\over k_1}  {\mc{P}(k_1)}W(k_1R_1)j_0(k_1r) \int_0^\infty {dk_2\over k_2} {\mc{P}(k_2)}W(k_2R_2)\int_{-1}^{1}d\mu W(kR) {\mc{A}(k)\over \mc{A}(k_1)\mc{A}(k_2)},\ee
where $\mu$ is the cosine of the angle between $\mb{k_1}$ and $\mb{k_2}$, and $k=\sqrt{k_1^2+k_2^2+2k_1k_2\mu}.$

With these definitions, Valageas showed via a saddle-point expansion that the bias for objects mass $M$ is given by
\be b^2(M,r) ={1\over {\sigma}^2_{0,0}(r)}\bkts{(1+\delta_R(s))e^{\Delta(s)}-1},\lab{valabias}\ee
where 
\be \Delta(s)&=& {{\sigma}^2_{R,R}(s)\delta_c^2\over u {\sigma}^2_R } + {2\delta_c^3\over u^3}\bkts{f_{R;RR}(s)+2g_{R;RR}(s)+\bkt{1- {u^3\over {\sigma}^6_R}}f_{R;RR}(0)},\\
u&=&{\sigma}^2_R +{\sigma}^2_{R,R}(s).\ee

\begin{figure}
\centering
\includegraphics[width= 6.2cm, angle = -90]{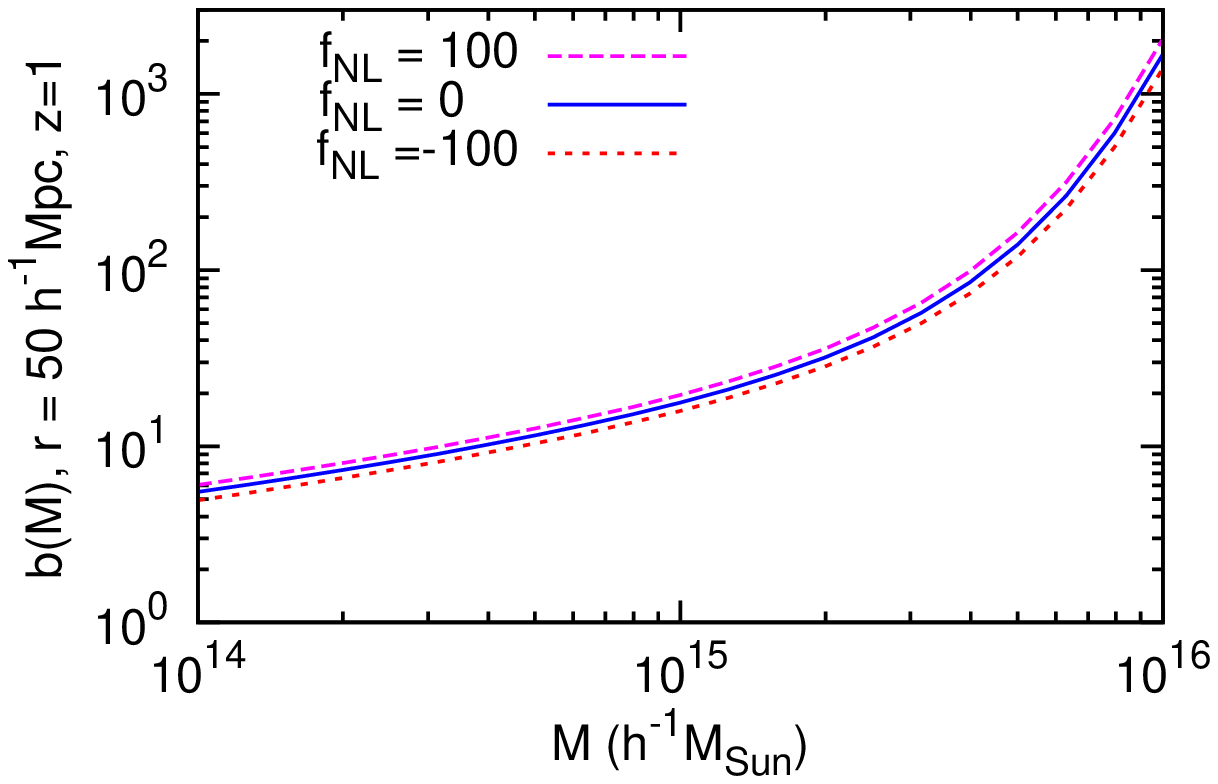}
\includegraphics[width= 6.2cm, angle = -90]{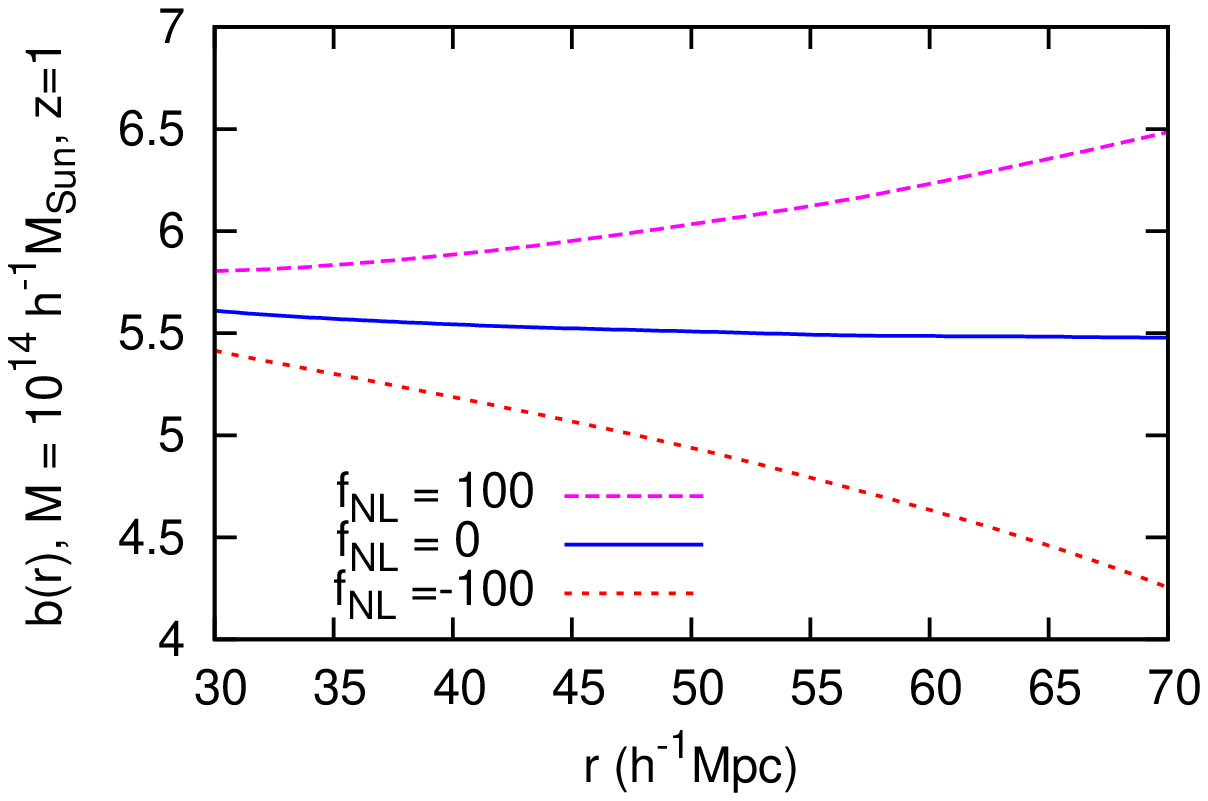}
\caption{The effect of $\fnl$ on the real-space bias, $b$, at $z=1$. In the panel on the left, the bias is shown as a function of smoothing mass-scale, $M$, with separation length $r=50h^{-1}$Mpc. The other panel shows the effect of varying $r$ with $M=10^{14} h^{-1}M_\sun$, illustrating the scale-dependence of the bias when $\fnl=\pm100$.}
\label{figbias}

\end{figure}

Figure \ref{figbias} shows the real space bias for a range of $M$ and $r$. Keeping $r$ fixed and varying $M$ (panel on the left), non-Gaussianity shifts $b(M)$ up or down (depending on the sign of $\fnl$). On the other hand, keeping $M$ fixed and varying $r$, we see how nonzero $\fnl$ introduces a scale dependence on $b(r)$ ($b(r)$ is roughly constant on large scale if $\fnl=0$). This scale-dependence is similar to that seen in \cite{me9} for $\gnl$.

In the limit of large separation length, the non-Gaussian bias $b(r)$ (with $M$ fixed) follows a simple scaling relation. Given $\fnl$, we can define the deviation from the Gaussian bias as $\Delta b\equiv b(\fnl)-b(\fnl=0)$. For $r\gg R$, it was shown that \cite{valageasA}
\be \Delta b\propto \fnl b(\fnl=0)\bkts{r\over h^{-1}\mbox{Mpc}}^2,\ee so that the overall scaling is $\Delta b\sim r^2$, since $b(\fnl=0)$ is approximately constant for large $r$. 

\begin{figure}
\centering
\includegraphics[width= 6.2cm, angle = -90]{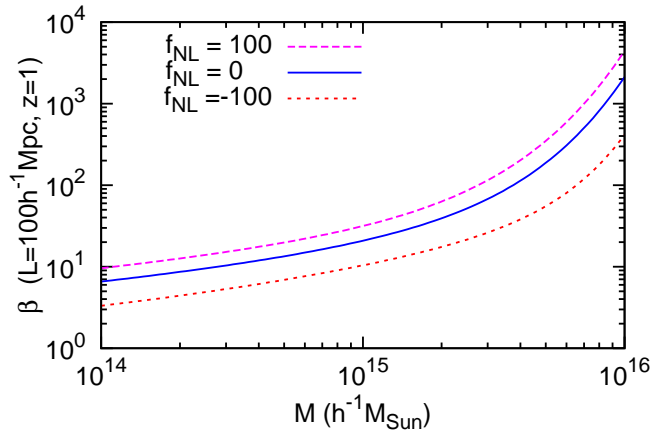}
\caption{The effective bias, $\beta$ (defined by Eq. \ref{beeta}) associated with objects of mass greater than $M$ in a spherical volume of radius $L=100h^{-1}$Mpc. Compared with figure \ref{figbias}, the effect of non-Gaussianity on $\beta$ is much more apparent.}
\label{figbeta}

\end{figure}

In later work, it will be necessary to define the effective bias associated with a comoving volume. For a spherical region of comoving radius $L$, we define such a bias as  
\be {b}_L(M) ={1\over V^2}\int_V \mb{dx}_1\int_V \mb{dx}_2 \ff b(M,|\mb{x}_1-\mb{x}_2|),\lab{b1}.\ee
It will also be useful to define the effective bias for objects of mass $>M$.
\be b(>M,r)\equiv{1 \over  n(>M)} \int_M^\infty {b}(m,r){dn\over dm}dm.\lab{b2}\ee 
Combining the averaging processes \re{b1}-\re{b2}, we can define
\be \beta(L,M) \equiv b_L(>M),\lab{beta}\ee
which, as shown in the appendix, simplifies in the limit $r\gg R(M)$ to
\be  \beta(L,M)\approx F(L)G(M),\lab{beeta}\ee
where 
\be F(L)&=&1+{6\over 5}K(z)\fnl \bkts{L\over h^{-1}\mbox{Mpc}}^2,\lab{fl}\\
G(M) &=& {1\over n(>M)}\int_M^\infty b(m, \fnl=0) {dn\over dm}\ff dm.\lab{gm}\ee
An example of the effective bias, $\beta$, with $L=100h^{-1}Mpc$ is shown in figure \ref{figbeta}, in which we set $z=1$ and use the Tinker mass function for $dn/dM$, although using a different mass function only results in small differences. Comparing this graph with that of $b(M)$ (left panel of figure \ref{figbias}), we see that whilst $\beta$ retains the overall shape of the curves, it is clearly more sensitive to non-Gaussianity. This quantity will be especially useful in the next section in which we consider the clustering of massive objects within a specified volume in the presence of non-Gaussianity.


We conclude this section with a brief comparison between the non-Gaussian imprints in the bias and in the cluster counts. Because the non-Gaussian imprint on the clustering of biased objects is significant only on large scales, surveys covering a large volume ($\sim \mc{O}(10)$ Gpc$^3$) will be required. Preliminary forecasts have shown good prospects of achieving $\fnl\sim\mc{O}(1)$ constraints from measurements of the bias with upcoming surveys such as DES\footnote{www.darkenergysurvey.org}, Euclid\footnote{http://sci.esa.int/euclid} and LSST\footnote{www.lsst.org} \cite{fedeli,carbone}. On the other hand, cluster number counts, whilst not requiring a large-volume survey, are almost completely insensitive to the shape of non-Gaussianity \cite{loverde,wagner0}. The bias probes correlation between scales and is therefore sensitive to the shape of non-Gaussianity, particularly the local shape, whereas the equilateral shape shows up only weakly in the bias \cite{verde}. This suggests that a combination of these probes will be required to constrain both the amplitude and the shape of non-Gaussianity.

\section{Extreme-Value Distributions}

In this section, we present the calculation of the distribution\footnote{We use the word `distribution' in the strict sense, referring to the cumulative  distribution and not the pdf.} of extreme-mass clusters. The necessary ingredients are the non-Gaussian number density and real-space bias calculated in the previous sections.

\subsection{Distribution function}

White \cite{white} derived the following expression for the cumulative probability that a region of volume $V$ contains no object of mass $M$ and above
\be P(M)= \exp\bkts{\sum_{k=1}^\infty {(-n(>m))^{k}\over k!} \bkt{\prod_{i=1}^{k}{\int_V \mb{dx_i}}}\, \xi\super{pk}_{k}(\mb{x_1},\mb{x_2},\ldots\mb{x_k})},\lab{white}\ee
where $\xi\super{pk}_1\equiv1$, $n(>M)$ is given by \re{en} and $\xi_k\super{pk}$ is the $k$-point correlation function of density peaks in $V$ associated with halos of mass $>M$. As in \cite{davis,colombi}, we shall at times refer to $V$ as a `patch'. If we take the patch to be a sphere of comoving radius $L$, the volume-averaged correlation then simplifies to the cumulant (connected moment) smoothed by a top-hat window of radius $L$ as follows
\be\bkt{\prod_{i=1}^{k}{\int_V {\mb{dx_i}\over V}}}\, \xi_{k}\super{pk}(\mb{x_1},\mb{x_2},\ldots\mb{x_k})&=&\bkta{\delta\sub{pk}^k}_c(L)\nn\\
&=& (\sigma\super{pk}(L))^{2k-2}S_k\super{pk}(L)\nn\\
&=& (\beta(L,M)\sigma_L)^{2k-2}S_k\super{pk}(L)\ee
where $\beta$ is given by \ref{beta}. 
The cumulants for density peaks have been calculated in the context of hierarchical structure formation with Gaussian initial condition \cite{manera, juszkiewicz, cooraysheth, bernardeau}. In the presence of non-Gaussianity, however, the perturbation theory required to calculate the cumulants for density peaks becomes much more complex (see \eg \cite{giannantonio,matsubara,bernardeau}). To make analytic progress, we shall consider only $k$ up to $3$ in the sum \re{white}. The terms $k=1$ and $2$ correspond to  well-known results previously found in \cite{colombi,sheth}, namely
\be -nV +{1\over2}(n V \beta \sigma_L )^2,\ee
where we have used the reduced cumulant $S_2=1$.
Given a weakly non-Gaussian initial condition, the third cumulants for density peaks are expected to be dominated by nonlinear gravitational effects, since primordial non-vanishing cumulants decay at the rate $S_k/D^{k-2}(z)$ \cite{fryscherrer}. 
Neglecting these effects, we can use the expression for the cumulant of the lognormal distribution \cite{kayo}
\be S_3\super{pk}(L) = 3+\sigma^2_L,\ee
which was found to be in fair agreement with $N-$body simulations of non-Gaussian models with $|\fnl|$ as large as $1000$ (at least in the quasi-linear regime with $\sigma\simeq1$) \cite{grossi}. This approximation is sufficient for the range of cluster masses ($\geq10^{14}h^{-1}M_\sun$) examined in this work.


Collecting these results, we find the extreme-value distribution
\be \ln P(M)&\approx& -X + {1 \over2}X^2Y^2-{1 \over6}X^3Y^4S_3\super{pk}(L),\lab{logp}\nn\\
\mbox{where}\quad  X&\equiv& n(>M)V, \quad Y \equiv \beta(L,M)\sigma_L.\ee

Setting \re{logp} equal $\log{(1/2)}$ (\iee the median value of $M\sub{max}$) gives an estimate of the modal value of $M\sub{max}$, at least in weakly non-Gaussian distributions (see \cite{sheth} for the Gaussian case). However, the equation is non-linear in $M\sub{max}$ and the approximate $M\sub{max}$ dependences in these terms are not intuitive. Instead, we look for the peak in the  derivative of \re{logp}, \iee the probability density function. Nevertheless, the shape of the distribution function holds valuable statistical information to which we shall return when we consider the extremal-type distributions in Appendix B.

\subsection{PDF of extreme-mass objects}

We can obtain the probability density function (pdf) for the most massive objects expected in a volume by differentiating the distribution function \re{logp} with respect to $M$, noting that the only dependence on $M$ is in the number density, $n$, and the bias, $b$. The result is


\be p(M)= {dP\over dM}=VP(M)\Bigg[-{dn\over dM}\bkt{-1+nV\beta^2\sigma_L^2-{1\over2}n^2V^2\beta^4\sigma^4_LS_3(L)}+\ff n^2V\beta\sigma_L^2{d\beta\over dM}\bkt{1-{2\over3}nV\beta^2\sigma^2_LS_3(L)}\Bigg],\lab{pdf}\ee
where $dn/dM$ is given by \re{dndm} (note the subtlety that $dn(>M)/dM=-dn/dM$). Here, we see explicitly that the pdf of extreme-mass objects not only depends on the bias, but also on its mass variation, $d\beta/dM$.


\subsection{$\fnl$ and extreme-value pdf}\lab{extremesubsec}


\begin{figure}
\centering

\hskip -1 cm\includegraphics[width= 4.8cm, angle = -90]{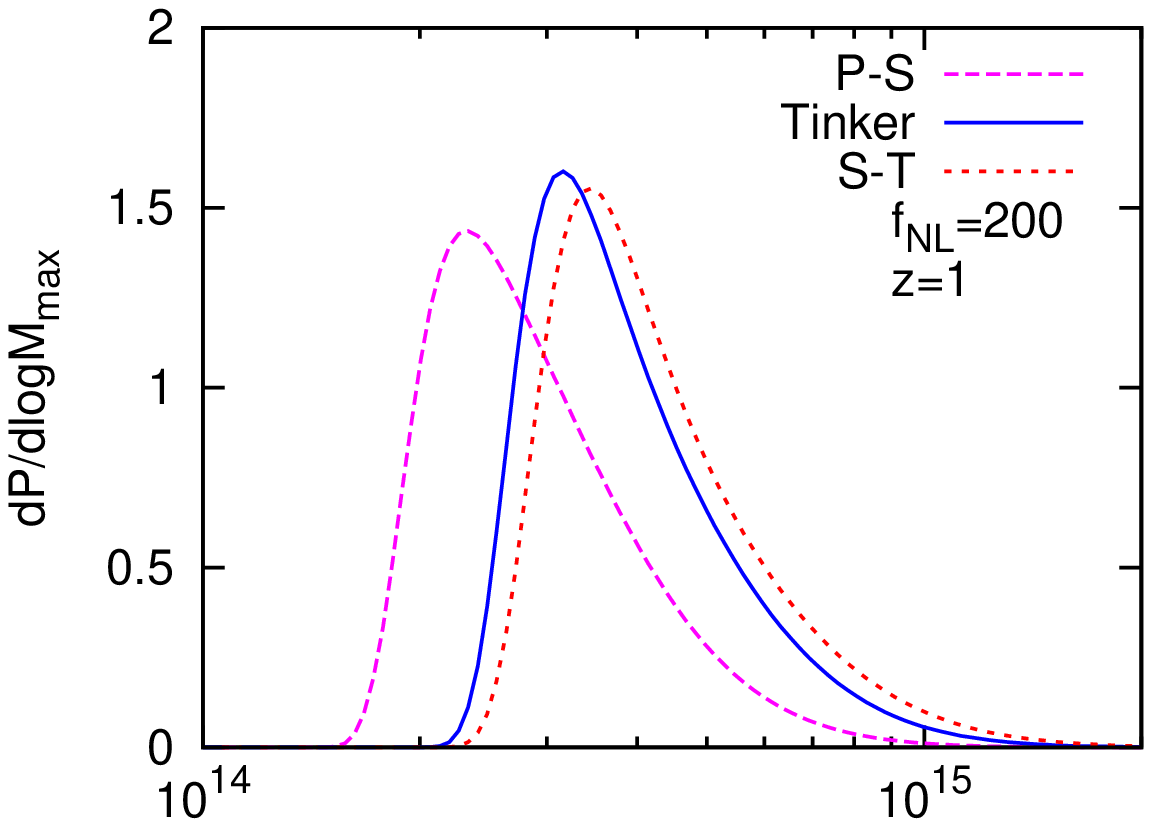}\hskip -1cm \includegraphics[width= 4.8cm, angle = -90]{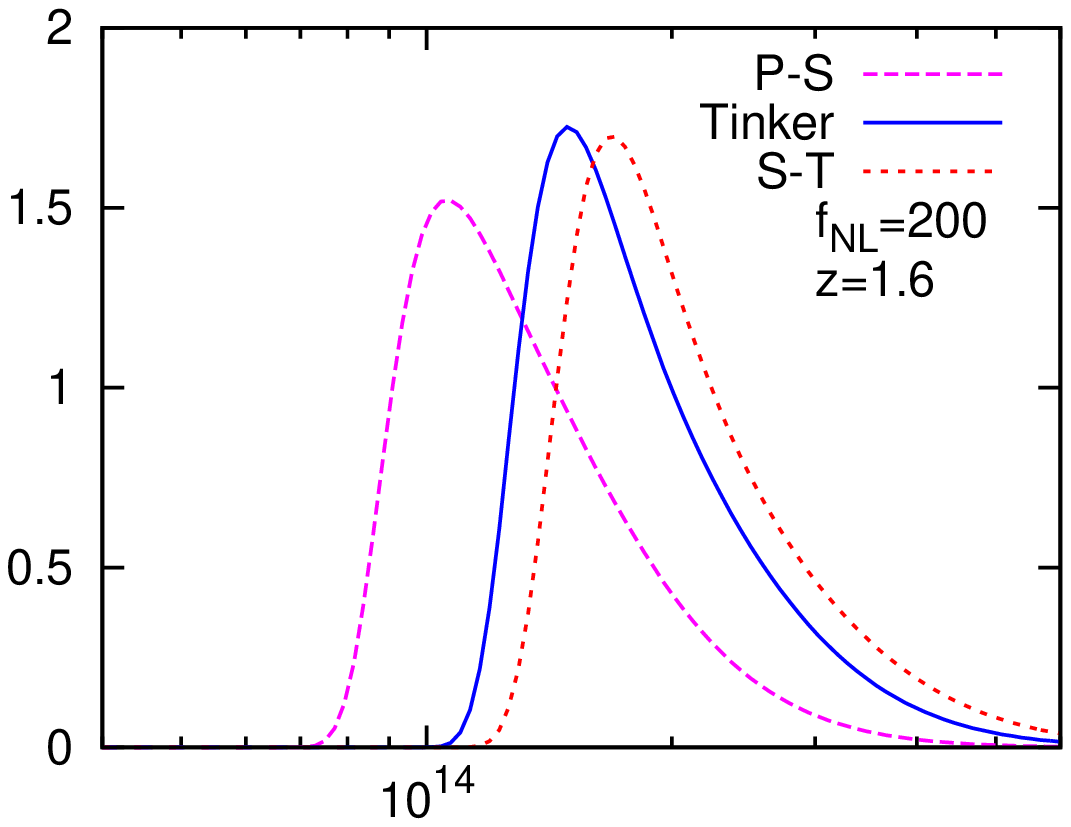}\hskip -1cm
\includegraphics[width= 4.8cm, angle = -90]{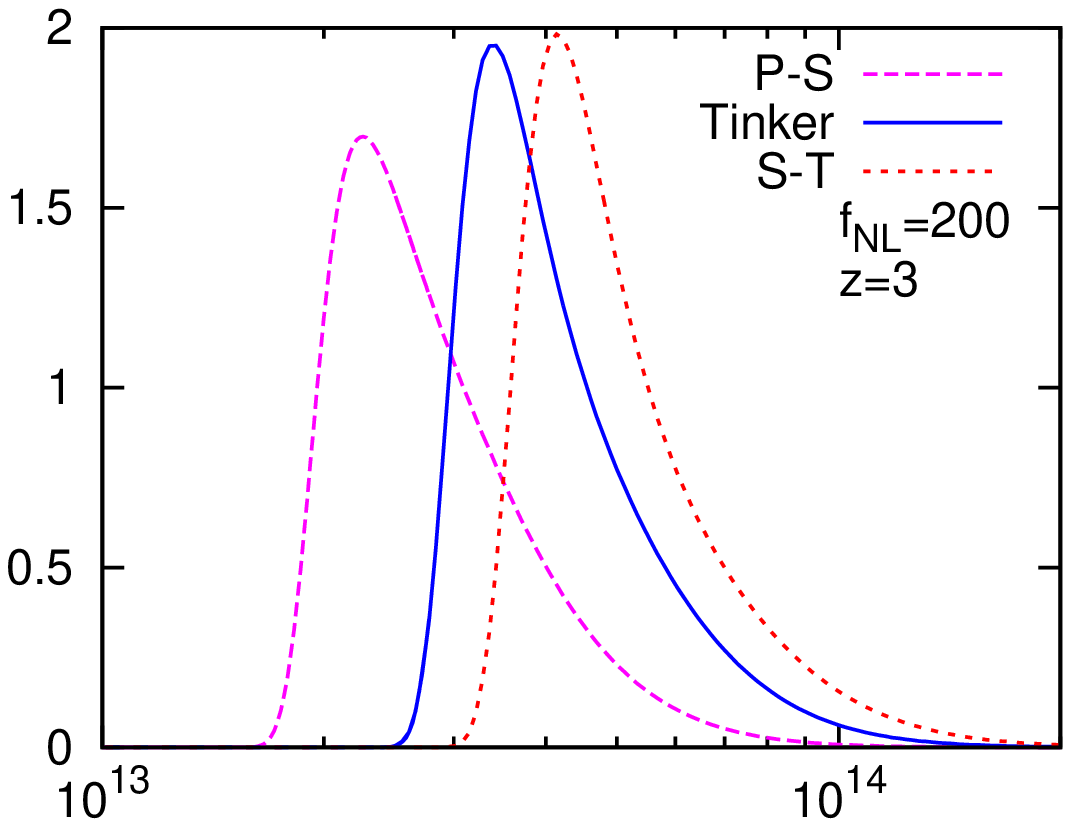}

\vskip -0.3 cm

\hskip -1 cm\includegraphics[width= 4.8cm, angle = -90]{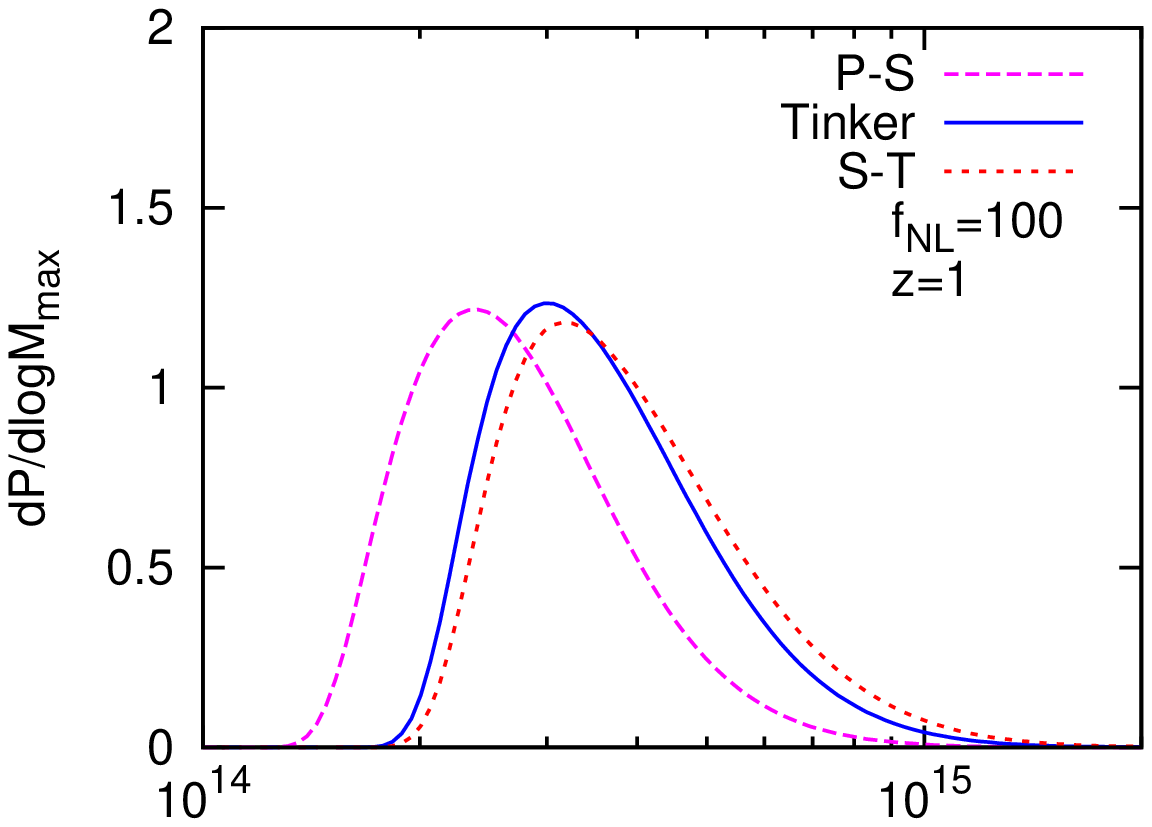}\hskip -1cm \includegraphics[width= 4.8cm, angle = -90]{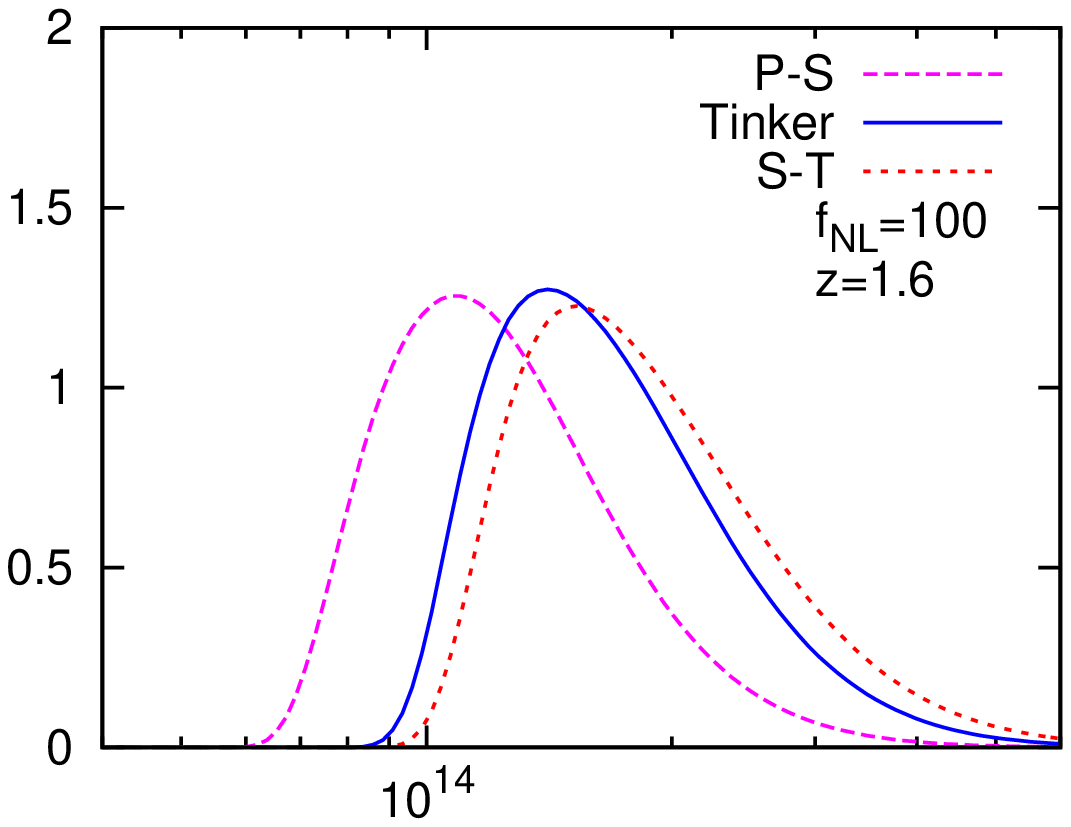}\hskip -1cm
\includegraphics[width= 4.8cm, angle = -90]{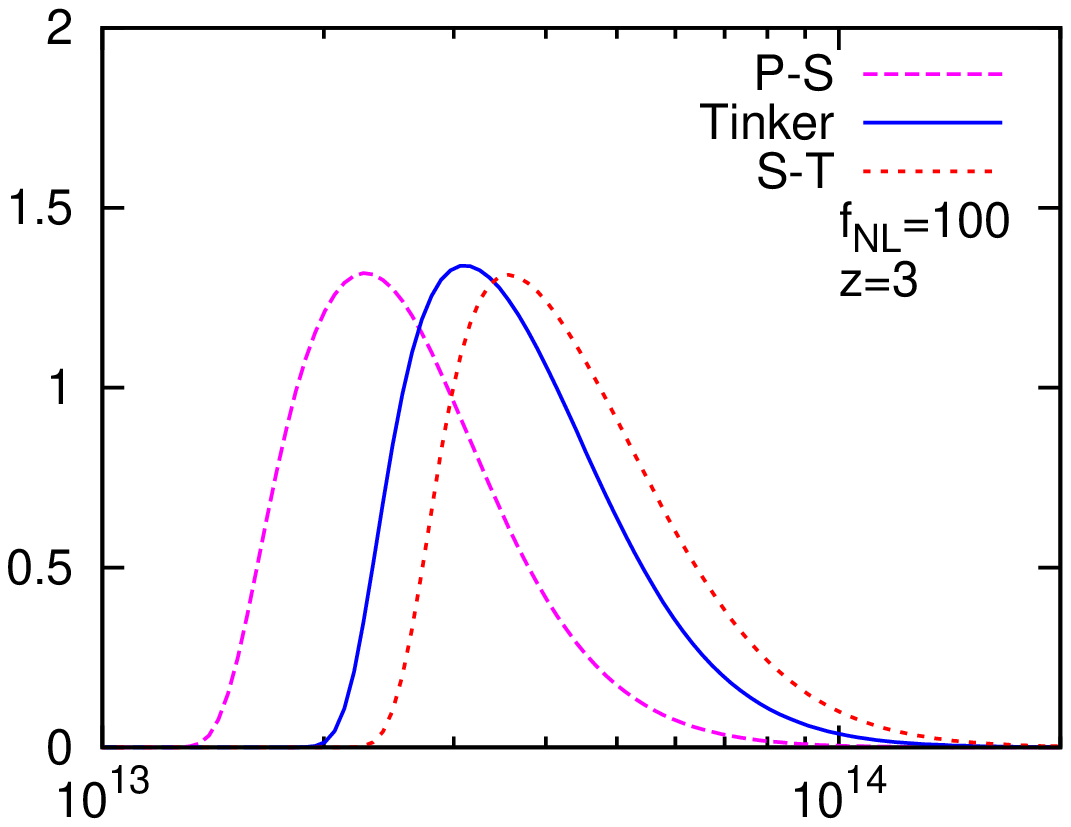}

\vskip -0.3 cm

\hskip -1 cm\includegraphics[width= 4.8cm, angle = -90]{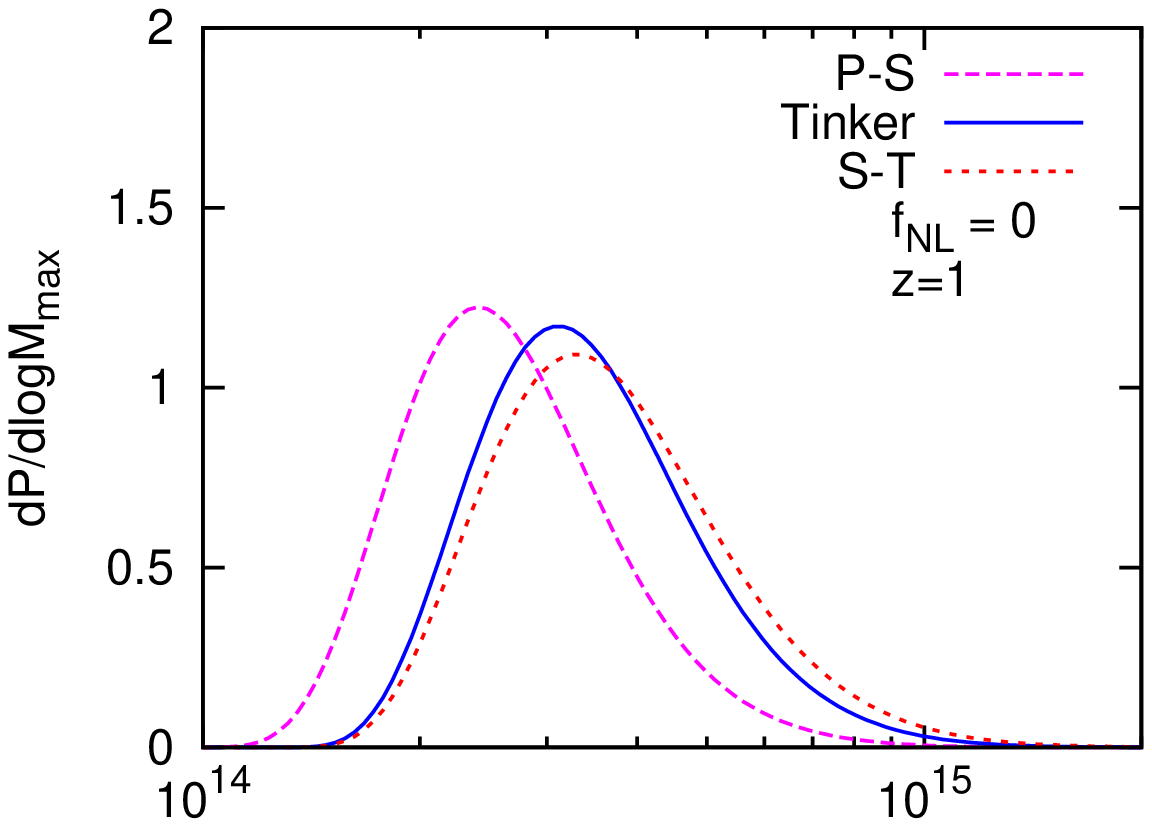}\hskip -1cm\includegraphics[width= 4.8cm, angle = -90]{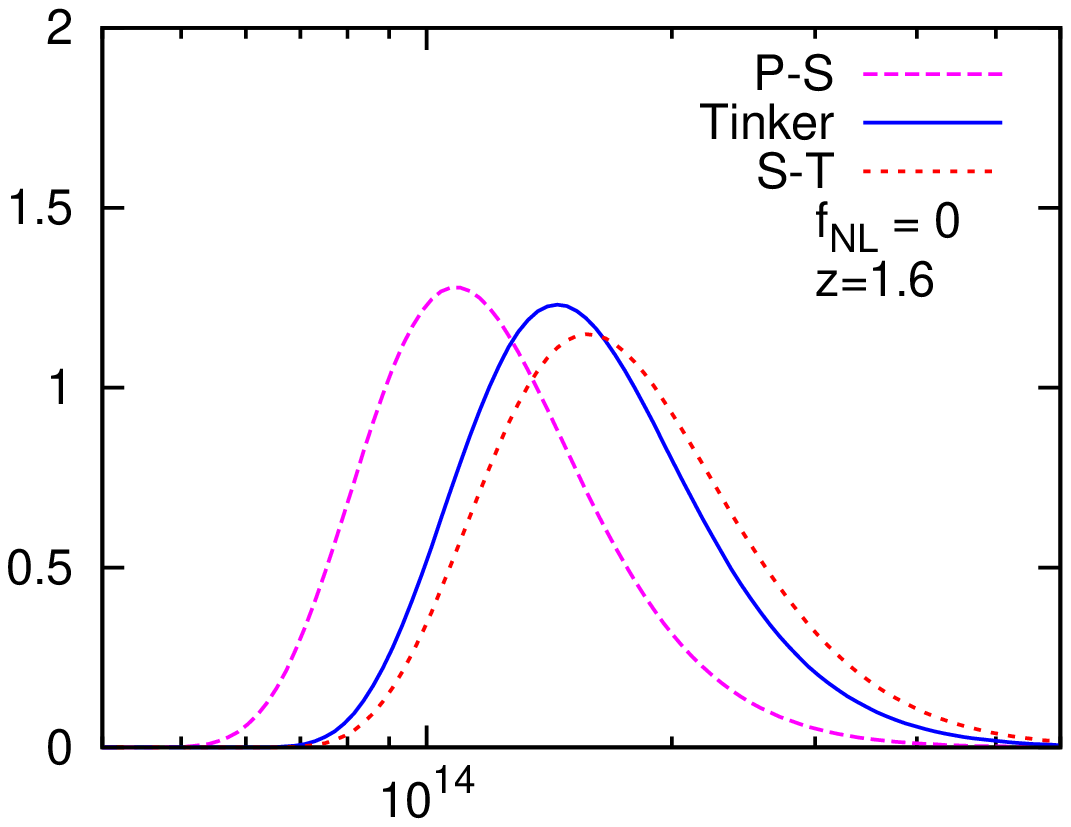}\hskip -1cm
\includegraphics[width= 4.8cm, angle = -90]{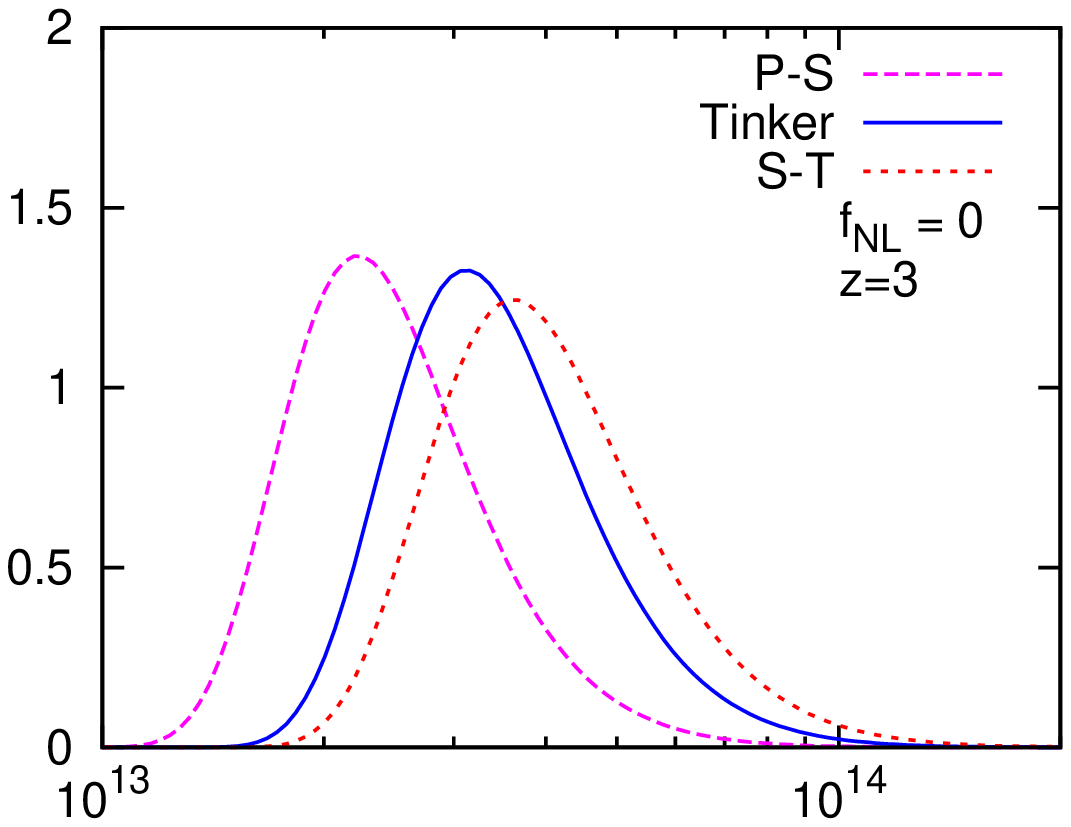}

\small{$M\sub{max} \ff(h^{-1}M_{\sun})$}
\caption{The probability density function of the maximum mass, $M\sub{max}$, of objects in a spherical volume of radius $L=100h^{-1}$ Mpc. In each panel, the mass functions used are Press-Schechter (dashed/magenta), Tinker (solid/blue) and Sheth-Tormen (dotted/red). \ii{Top row:} $\fnl=200$ with $z=1,1.6$ and 3 (from left to right). The pdf at the same redshifts are shown for $\fnl=100$ (middle row) and $\fnl=0$ (bottom row). The non-Gaussian effects are most visible in the third column in which the peaks can be seen to move to higher $M\sub{max}$ with increasing $\fnl$.
}
\label{figbig}

\end{figure} 


The main results of this paper are shown in figure \ref{figbig}. The panels show  the probability density function \re{pdf} for the three mass functions at redshift $z=1,1.6$ and 3 (corresponding to the left, middle and right column) with $\fnl=200,100$ and 0 (top, middle and bottom row respectively). The survey volume is taken to be a sphere of radius $100h^{-1}$Mpc. To display the correct scaling on the horizontal log scale, we plot $dP/d\log M\sub{max}$ on the vertical axis  whilst the actual value of the pdf is $dP/dM\sub{max}$. From these graphs, we make the following observations:
\bit
\item[(a)] Going from the bottom row to the top, we see that increasing $\fnl$ increases the height of the pdf whilst positively skewing it (\iee lifting the positive tail). This has the effect of increasing the mass of the most probable extreme objects in a given volume. 
\item[(b)] Going from the first column to the third, we see that at higher redshifts, the pdfs are more peaked and the peaks are located at lower $M\sub{max}$. 
\item[(c)] The Sheth-Tormen mass function predicts the largest mass of extreme objects, followed by the Tinker and the Press-Schechter mass functions. This is a consequence of their predicted number densities as seen in figure \ref{fign}.
\item[(d)] The differences between the mass-functions become much more pronounced at high redshifts. In the third column, we see a clear separation of the peaks for different mass functions, with non-Gaussianity further enhancing the differences. 

\begin{figure}

\centering




\includegraphics[width= 5cm, angle = -90]{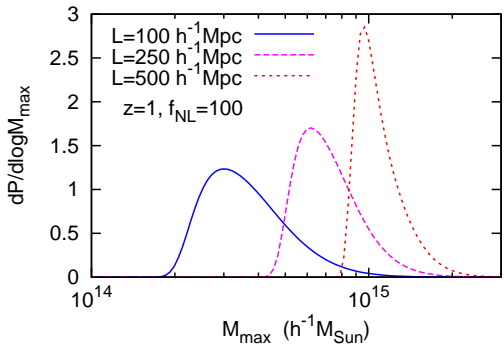}

\caption{The extreme-value pdf for patch sizes $L=100h^{-1}$Mpc (solid/blue), $250h^{-1}$Mpc (dashed/magenta) and $500h^{-1}$Mpc (dotted/red). The Tinker mass function is used, with $z=1$, and $\fnl=100$. We see that the location of the peak is clearly very sensitive to changes in $L$.}
\label{varyL}
\end{figure}

\item [(e)] In figure \ref{varyL}, we show the effect of varying the patch radius, $L$ from 100 to 500 $h^{-1}$Mpc (with $z=1$ and $\fnl=100$). By increasing $L$, the peak of the pdf shifts significantly to higher $M\sub{max}$. The pdf also  becomes more peaked with increasing $L$. This is simply due to the fact that as the sample size, $L$, approaches the population size, repeating the sampling will yield almost identical maxima in the samples.

\eit

\begin{figure}

\centering

\hskip -0.8 cm\includegraphics[width= 4.6cm, angle = -90]{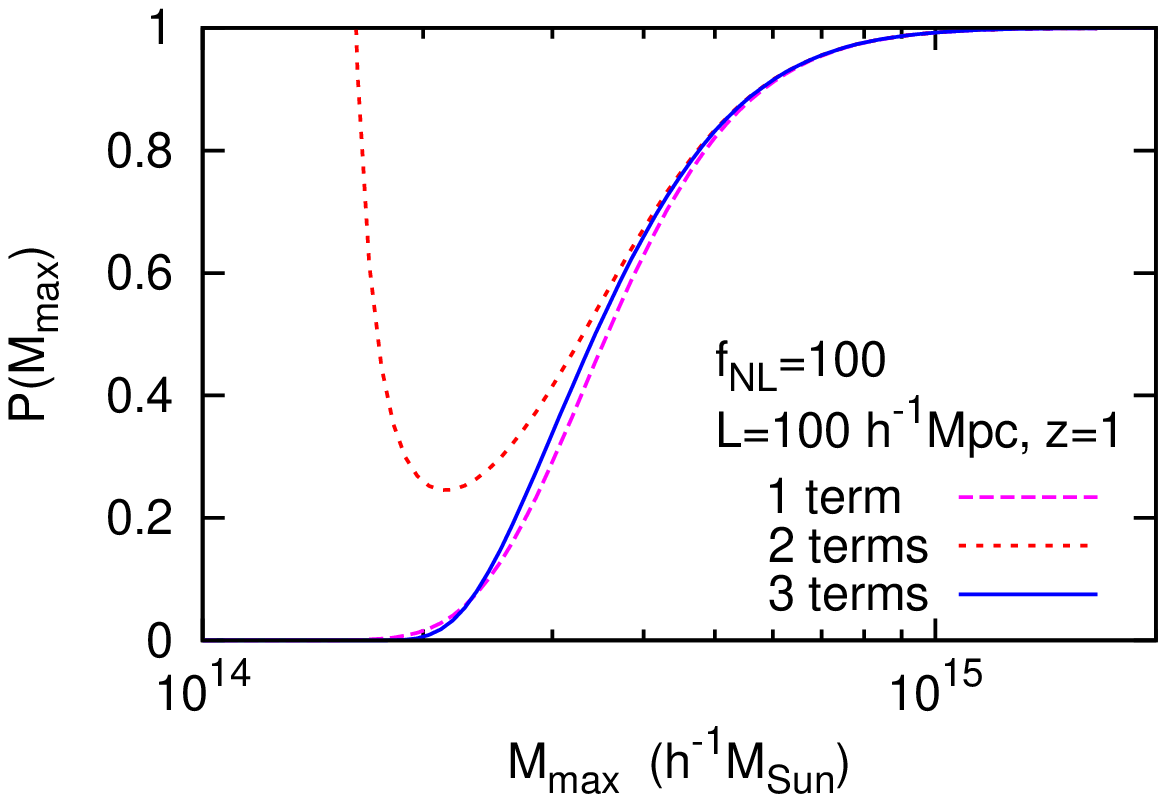}
\hskip -0.8 cm\includegraphics[width= 4.6cm, angle = -90]{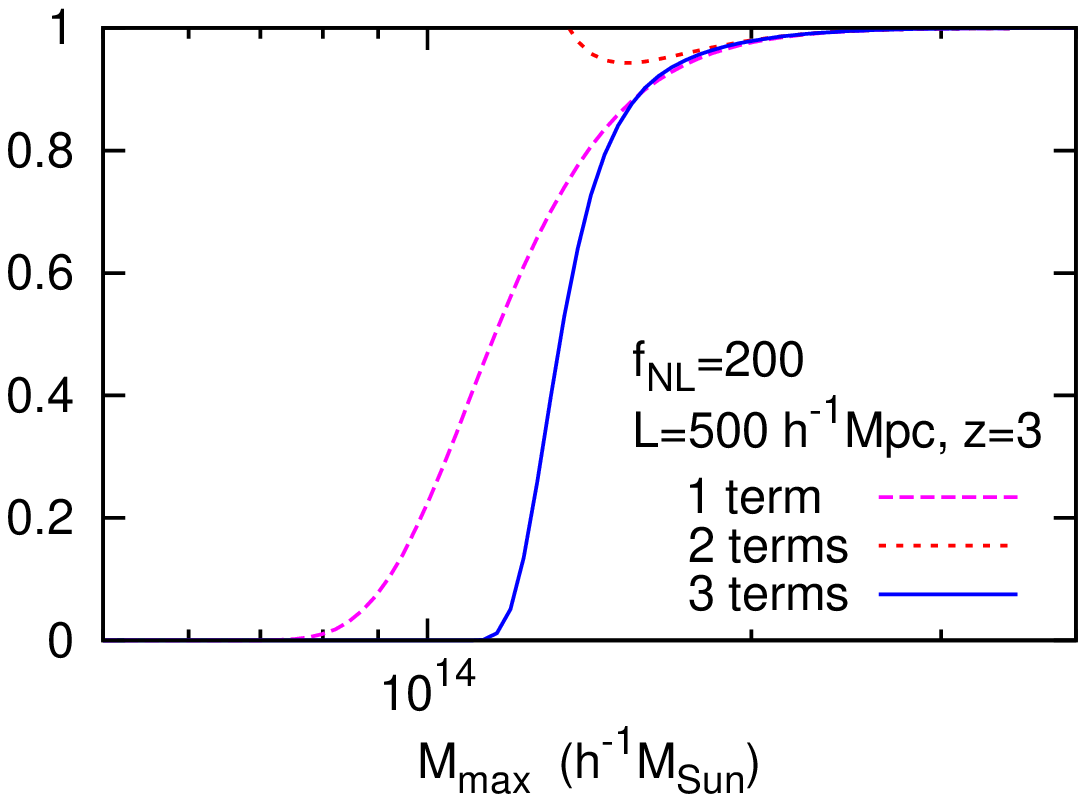}
\hskip -0.8 cm\includegraphics[width= 4.6cm, angle = -90]{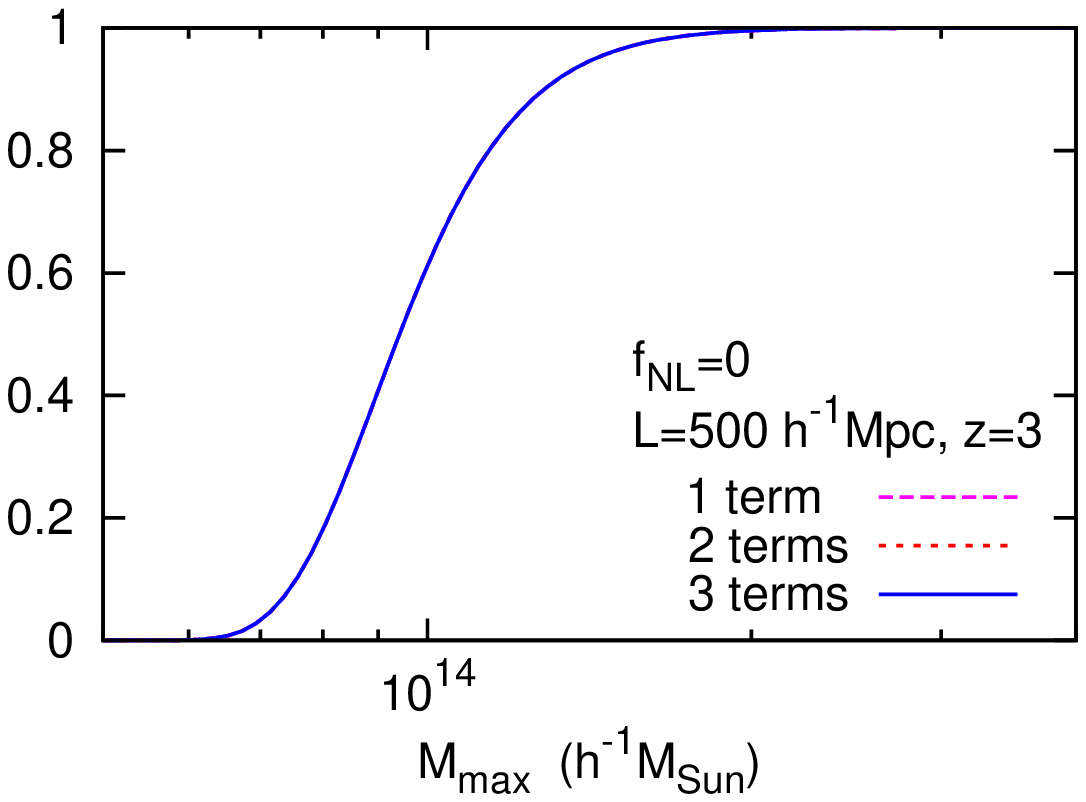}

\caption{The contributions of the 3 terms in \re{logp} towards the extreme-value distributions, using 3 sets of parameters \ii{Left:} (A) $\fnl=100$, $L=100h^{-1}$Mpc, $z=1$. \ii{Middle:} (B) $\fnl=200$, $L=500h^{-1}$Mpc, $z=3$. \ii{Right:} (C) $\fnl=0$, $L=500h^{-1}$Mpc, $z=3$. In case (B), the Poisson approximation (`1-term') is clearly inaccurate.}
\label{figrelative}
\end{figure}

Finally, we investigate the relative importance of the three terms on the right-hand side of equation \re{logp}. We consider the extreme-value distributions in  following cases
\bit
\item[(A)] $\fnl=100$, $L = 100 h^{-1}$Mpc, $z=1$,
\item[(B)] $\fnl=200$, $L = 500 h^{-1}$Mpc, $z=3$,
\item[(C)] $\fnl=0$, $L = 500 h^{-1}$Mpc, $z=3$.
\eit
Figure \ref{figrelative} shows the resulting distributions when one, two or three   terms on the right-hand side of \re{logp} are taken into account (using the Tinker mass function). In case (A), we see that the Poisson approximation (keeping only the first term in \re{logp}) is fairly close to the 3-term result. Generally, this holds as long as the non-Gaussian effects are small (\iee at small volume, low redshift). However, in case (B), we see that the Poisson approximation underestimates the extreme cluster masses. In this case the peak of the pdf, though merely shifted by $\lesssim 10 \%$, is much narrower and rises to a higher maximum value (as can be estimated by eye from the slope around $P=0.5$). In both cases, however, it is inconsistent to include the bias but ignore the third term (representing the skewness of the halo distribution in the sample). This is shown in the upturns of the 2-term distribution functions. In case (C), where $\fnl=0$, the 2nd and 3rd correction terms are negligible and the Poisson approximation is very good indeed.
 
Strictly speaking, the calculations here are valid only for $L\gg r$. In actual applications, we will be interested in the cases where $L=\mc{O}$(a few Gpc). In such cases, the redshift variation within the patch must be taken into account (as emphasised in \cite{davis}). This requires replacing the number density, bias and cumulants by their average within a comoving volume. We shall demonstrate this in the next section.

\section{Applications}

\subsection{A massive cluster at $z = 1.579$ : a problem for $\Lambda$CDM?}

Santos \etal \cite{santos} recently reported the discovery of a cluster XMMUJ0044.0-2-33 (hereafter XMMUJ0044) at $z=1.579$, detected in the X-ray data of the XMM-Newton telescope and later followed up spectroscopically. The cluster mass was estimated to be $\sim3.5-5\times10^{14}M_\sun$, far greater than the previous X-ray cluster of mass $5.7\times10^{13}$ at $z=1.62$ reported by Tanaka, Finoguenov and Ueda \cite{tanaka}. We shall now use extreme-value statistics to study the probability of finding XMMUJ0044 as the maximum-mass cluster. In particular, we ask, is the existence of XMMUJ0044 consistent with $\fnl=0$? 
 
\subsection{Eddington bias}

We take the mass of XMMUJ0044 to be\footnote{The mass of XMMUJ0044 published in \cite{santos} is $(4.25\pm0.75)\times 10^{14} M_\sun$, given with respect to the critical density. Assuming NFW cluster profile, Waizmann \etal \cite{waizmann} showed that, with respect to the mean background density, the value increases to $(4.46\pm0.79)\times 10^{14} M_\sun$. We thank J.C. Waizmann for bringing this point to our attention.} $M\sub{obs}=(4.46\pm0.79)\times10^{14}M_\sun$ (=$(3.12\pm0.55)\times10^{14}h^{-1}M_\sun$) and ask: what is the probability that this cluster is the most massive one in the redshift range $1.579\lesssim z \lesssim 2.2$? Here the redshift upper bound is consistent with the highest redshift probed by the XMM survey.

The reported cluster mass must first be corrected for Eddington bias, which refers to the apparent boost in the cluster mass due to the fact that it is more likely for lower-mass objects to scatter to high luminosity than it is for rarer massive objects to scatter to lower luminosity. We account for this effect by the correction \cite{mortonson}
\be \ln M=\ln M\sub{obs}+{1\over2} \gamma \sigma\sub{lnM}^2,\ee
where $\sigma\sub{lnM}\sim0.3$ is the error estimated from the observation and $\gamma$ is the local slope of the mass function determined using the relation $dn/d\ln M \propto M^{\gamma}$, and therefore satisfies
\be \gamma = {d^2n\over du^2}\bigg{/}{dn\over du}, \quad u \equiv \ln M.\ee
The final masses with Eddington-bias corrections are listed in Table \ref{tabi}. The corrected mass depends on $n(>M)$ and hence it also depends on the mass function used. There is also a weak dependence on $\fnl$ (entering via the MVJ correction \re{mvj}). With $\fnl=100$, the changes in the corrected masses are less than a percent and it is reasonable to neglect this correction as long as $|\fnl|\lesssim\mc{O}(10^2)$. We quote the corrected mass with $\fnl=0$ in Table \ref{tabi}.

\begin{table}[b]
\centering
\caption{Observed and Eddington-corrected mass for the cluster XMMUJ0044, in units of $10^{14} h^{-1}M_\sun$.}
\begin{tabular}{ccccc}
\hline
Observed mass &\ff& & Eddington-corrected mass&  \\
$(10^{14} h^{-1}M_\sun)$ &\ff& Press-Schechter & Sheth-Tormen& Tinker \\
\hline
$3.12\pm0.55$&\ff & $2.48\pm0.40$ &$2.62\pm0.43$& $2.56\pm0.41$ 
\end{tabular}
\label{tabi}
\end{table}

\subsection{Redshift averaging}

The patch of interest is now a spherical shell whose thickness is determined by the redshift band $\Delta z$. To account for the redshift variation within the shell, we perform the following modifications to the variables $X$ and $Y$ in the distribution function \re{logp}.
\be X&=& \bkta{n(>M)}V  =\int_{\Delta z} dz\int_M^{\infty}dm\ff {dn\over dm}{dV\over dz}.\lab{zave}\\
Y &=&\beta\sub{shell}(M)\bkta{\sigma}.\ee
The bias averaged within the shell, $\beta\sub{shell}$, is derived in Appendix A (Eq. \ref{fshell}) and is given by 
\be\beta\sub{shell}= {L^3\bkta{F(L)}+\ell^3 \over L^3 -\ell^3}\bkta{G(M)}.\ee 
Here $L$ and $\ell$ are comoving lengths corresponding to redshifts $2.2$ and $1.579$ respectively. $\bkta{\sigma}$, $\bkta{F(L)}$ and $\bkta{G(M)}$ are the redshift averages of \re{vari}, \re{fl} and \re{gm}. We define the redshift average of a quantity, $Q(z)$, by
\be \bkta{Q}={1\over V}\int_{\Delta z} dz\ff Q(z){dV\over dz}.\ee





\subsection{Results}

We are interested in the probability that a cluster of mass in the range $[\bar{M}+\sigma,\bar{M}-\sigma]$ is the maximum-mass object observed in a survey with a given $f\sub{sky}$ and redshift range. Denoting this probability as $\Pi$, we can express it as the difference in the distribution function \re{logp} evaluated at $\bar{M}\pm\sigma$.
\be \Pi = P(\bar{M}+\sigma)- P(\bar{M}-\sigma).\lab{Pi}\ee
We take the mass range to be those in shown in Table \ref{tabi}. In addition to the choice of mass function explored in the previous sections, here we consider three other factors that affect $\Pi$.

\subsubsection{Dependence on $f\sub{sky}$.}

The XDCP survey covers a sky area of approximately 80 deg$^2$ \cite{santos}. However, the value of $f\sub{sky}$ appropriate for our calculation must take into account all previous surveys that have explored the redshift interval in other parts of the sky, regardless of whether a positive detection is reported. 

\begin{figure}
\centering
\hskip -0.5 cm \includegraphics[width= 6cm, angle = -90]{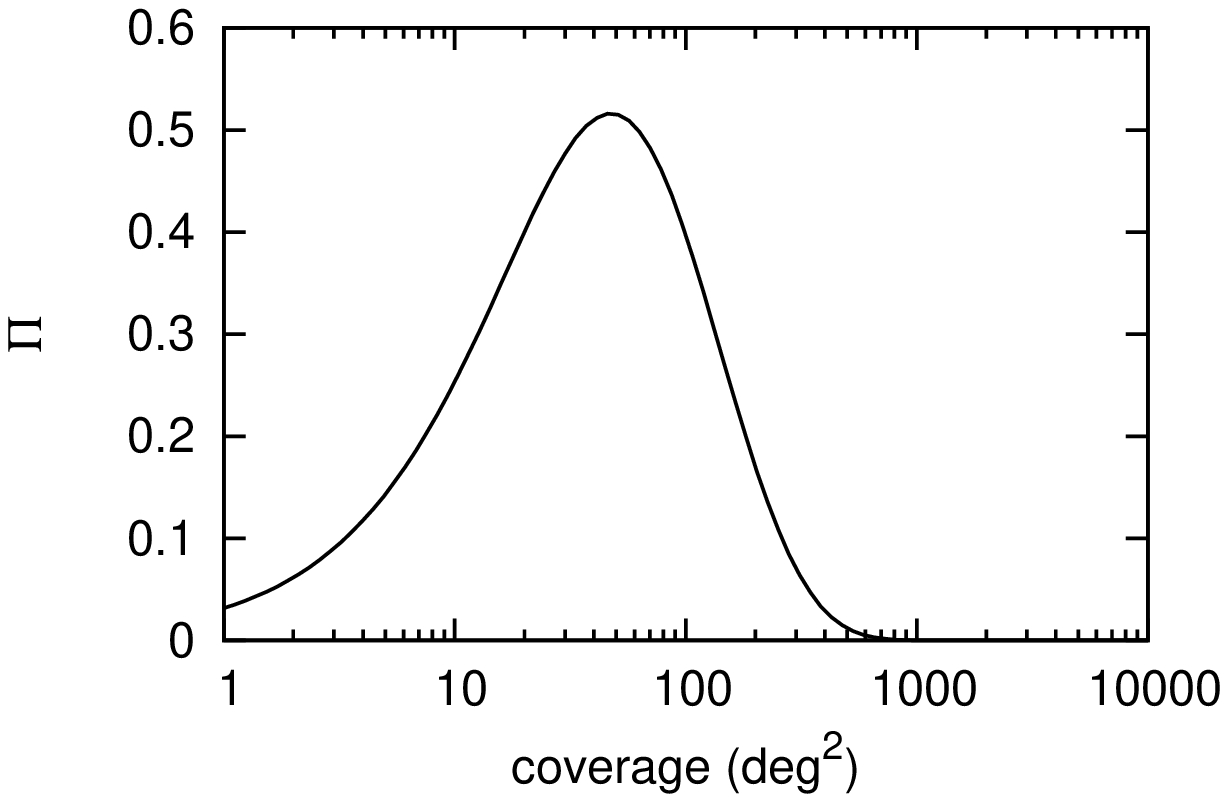}\ff\ff\ff
\hskip -0.5 cm \includegraphics[width= 6cm, angle = -90]{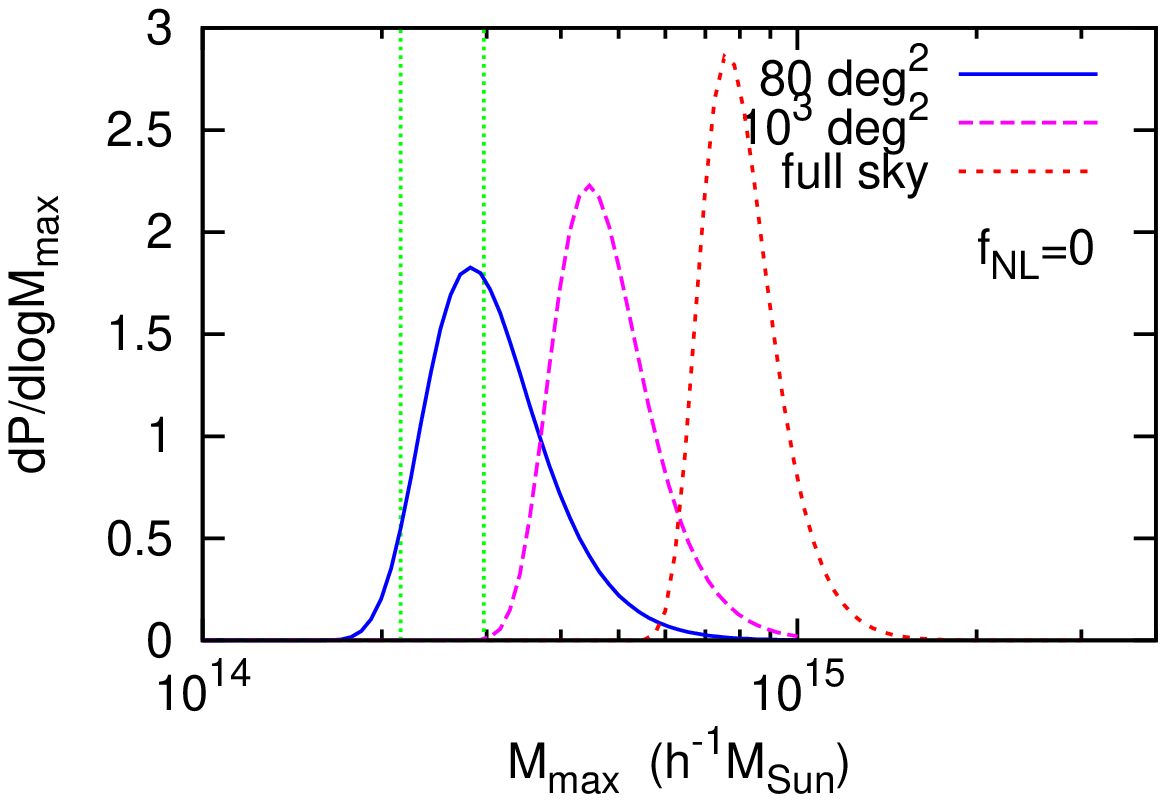}
\caption{\ii{Left:} The probability $\Pi$ (Eq. \ref{Pi}) that XMMUJ0044 is the most massive cluster in $1.579<z<2.2$ plotted as a function of sky coverage (in square degree). Using the Tinker mass function, the probability is maximised to $\approx50\%$ when $f\sub{sky}$ is of order 50 deg$^2$. \ii{Right:} The extreme-value pdf with coverages (from left to right) $80$ deg$^2$, $1000$ deg$^2$ and full sky (using the Tinker mass function and $\fnl=0$). Note from Table \ref{tabi} that the cluster lies at $2.56\pm0.41h^{-1}M_\sun$ (vertical dotted lines).}
\label{figsky}
\end{figure}

In figure \ref{figsky} (left panel), we show $\Pi$ as a function of survey area in square degree. Here $\Pi$ is calculated using the Tinker mass function and $\fnl=0$. We see that the observation of an extreme object such as XMMUJ0044 is most likely in a survey area of around 50 deg$^2$ (where $\Pi\simeq0.5$). At wider coverages, we expect the most probable extreme mass to be larger. In fact, taking $f\sub{sky}=1$ as the most conservative limit, we find the most likely extreme object in this redshift range to be a cluster of mass $\sim7-8\times10^{14}h^{-1}M_\sun$ (figure \ref{figsky}, right panel), well above the Eddington-corrected mass of XMMUJ0044. On the other hand, taking $\fsky=80$ deg$^2$, the most probable extreme-mass object is consistent with XMMUJ0044, as the peak of the extreme value pdf lies within the mass estimate (vertical contours in figure \ref{figsky}).

It is difficult to estimate what is the correct value of $f\sub{sky}$ is needed in such cases and, unfortunately, the question of whether non-Gaussianity is needed to explain the existence of certain clusters depends sensitively on the value of $\fsky$ assumed. Making the most conservative interpretation using $\fsky=1$ and  assuming that there are good prospects for discovering many more massive high-redshift clusters in the future, we conclude that XMMUJ0044 presents no tension with $\Lambda$CDM (see also the conclusions of \cite{mortonson,waizmann2}).

\subsubsection{Dependence on $\fnl$.}

\begin{figure}
\centering
\hskip -0.5 cm \includegraphics[width= 6cm, angle = -90]{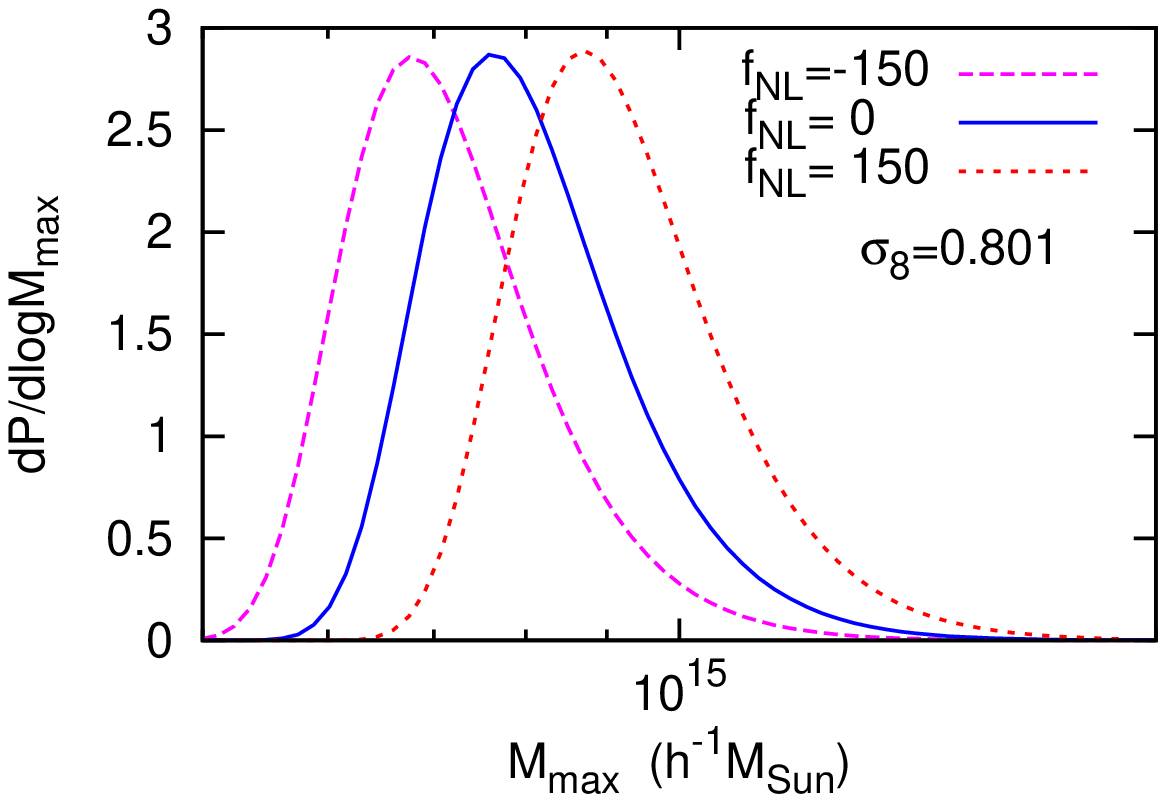}\ff\ff\ff
\hskip -0.5 cm \includegraphics[width= 6cm, angle = -90]{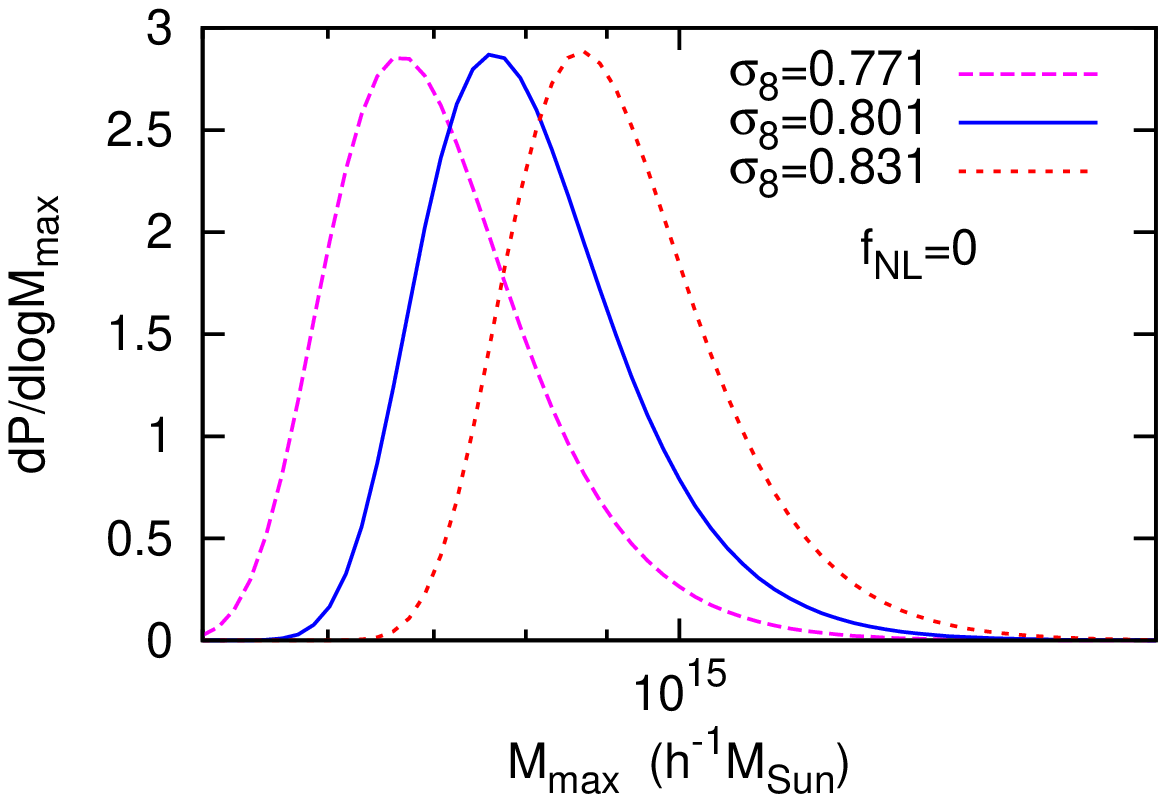}
\caption{The extreme-value pdf for objects in the range $1.579<z<2.2$  with a fiducial value for $\fsky= 1$, using the Tinker mass function. \ii{Left:} $\fnl$ is varied from $-150$ to $150$ with $\sigma_8=0.801$ \ii{Right:} $\sigma_8$ is varied in the range $\sigma_8=0.801\pm0.03$ \cite{komatsu} with $\fnl=0$. The degeneracy is discussed in the text.}
\label{figothers}
\end{figure}

In figure \ref{figothers} (left panel) we show the effect of introducing $\fnl=\pm150$ to the extreme-value pdf. Here we take a fiducial value of $\fsky=1$, and use the Tinker mass function. As expected, the pdf is shifted left or right depending on the sign of $\fnl$. 

\subsubsection{Dependence on $\sigma_8$}

The panel on the right of figure \ref{figothers} shows the equivalent effect when $\sigma_8$ is varied in the range $\sigma_8=0.801\pm0.03$ (taken from the WMAP 7-year constraint \cite{komatsu}), whilst $\fnl$ is fixed to $0$. The shift of the pdf towards more massive extreme objects follows from the fact that a greater $\sigma_8$ introduces a larger spread in the mass range of cluster in the survey volume.

The similarity between the two panels of figure \ref{figothers} reflects the well-known degeneracy between $\fnl$ and $\sigma_8$ \cite{robinson,ribeiro}. This degeneracy can be easily broken, for instance, by the constraint on the galaxy power spectrum (which is sensitive to $\sigma_8$ but not $\fnl$) or the CMB temperature anisotropies \cite{cayon}.

\mmm

In summary, the degenerate effects between $\fsky,\fnl$ and $\sigma_8$ shown here imply that it is very difficult to deduce information on non-Gaussianity from the extreme-value distributions alone. The most sensible approach is combine the results from several cluster surveys (to achieve $\fsky=1$) with probes of the CMB (to break the $\sigma_8$ degeneracy), assuming selection effects and error in the mass determination can be kept in check.

\subsection{The most massive object in the Universe}

\begin{figure}
\centering
\hskip -0.5 cm \includegraphics[width= 6cm, angle = -90]{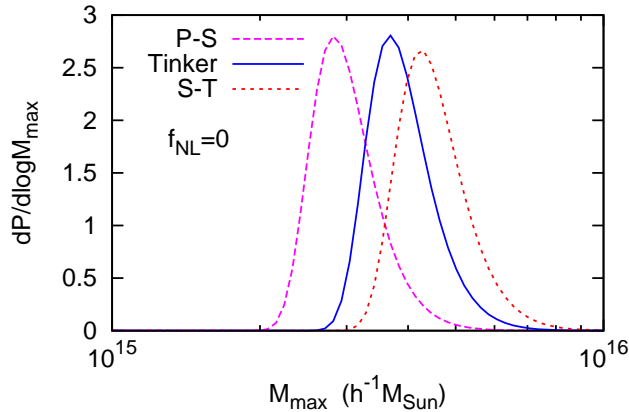}
\caption{The extreme-value pdf for the most massive object in the Universe (with $f\sub{sky}=1$ and $\fnl=0$) for various mass functions. }
\label{figmost}
\end{figure}

As a consistency check, we plot the extreme-value pdf for an extended redshift range $z>0$ and $\fsky=1$ in figure \ref{figmost}. This gives the extreme-value pdf for the most massive object in the Universe. With $\sigma_8=0.801$, We find this to be an object of mass $2-5\times 10^{15}h^{-1}M_\sun$, depending on the mass function. The result using the Tinker mass function is $M\sub{max}\approx 3.5\times 10^{15}h^{-1}M_\sun$ which agrees broadly with those reported by \cite{holz,davis}. The effect of $\fnl=100$ increases this value by by less than $10\%$.

\subsection{Extremal types}

Extreme-value distributions obey a limit theorem analogous to the Central Limit Theorem. This is the so-called \ii{Extremal Types Theorem}, which roughly states that extreme-value distributions converge to one of three possible types in the large-sample limit (see Appendix B). This beautiful theorem has found applications in areas such as meteorology, engineering and finance, where a large volume of data allows extreme-value statistics to be modelled by fitting only one or two parameters of an extremal type (analogous to fitting the mean and variance of the normal distribution) \cite{kotz,reiss}.

In cluster cosmology, the dearth of cluster data at present is not ideal for application of the limit theorem, although there have been attempts to apply it to simulated data (see \cite{davis,colombi}). The question of which extremal type extreme-mass clusters converge to remains unclear, although there is some evidence from simulations that $\fnl$ appears to play no role in the convergence \cite{mikelsons,harrison2}. We concur with this latter view and present the detail our investigation in Appendix B.

\section{Conclusion and Discussion}

In summary, we have investigated quantitatively how the statistics of extreme-mass clusters is affected by uncertainties in the mass function, non-Gaussian corrections of the mass function and bias, Eddington correction, $\fsky$, redshift, $\fnl$ and $\sigma_8$. More specifically,

\begin{enumerate}
\item We have presented a procedure to calculate the statistics of extreme-mass galaxy clusters in the presence of primordial non-Gaussianity parametrized by $\fnl$. Our main results are the expressions for the cumulative probability distribution \re{logp} and the probability density function \re{pdf} for the most massive object in a survey of a given sky coverage and redshift range. These expressions enable us to deduce the most probable extreme-mass cluster in a survey of a given specification. The effects of changing the mass function and varying the value of $\fnl$, survey volume and redshift are summarised in figures \ref{figbig} and \ref{varyL}.

\item Our correction terms for the extreme-value distribution (second and third terms on the right-hand side of \re{logp}) are significant when considering a large volume, high redshift or large non-Gaussianity (see figure \ref{figrelative}). For non-Gaussianity with $\fnl=\mc{O}(1)$, the first term of \re{logp} (Poisson approximation) suffices.

\item Next, we applied our formalism to investigate the extreme-value properties of cluster XMMUJ0044.0-2-33 ($M\sub{obs}\simeq 3.12\times10^{14}h^{-1}M_\sun$ at $z=1.579$). We find that the probability that the cluster is the most likely extreme-mass cluster in the survey depends sensitively on the assumed sky coverage, but is consistent with $\fnl=0$ (assuming $\sigma_8=0.801$). More conservatively, with $\fsky=1$, the most probable extreme-mass cluster is expected to be much larger and perhaps this will be confirmed by future X-ray cluster surveys.

\item We show that the effect of $\fnl$ in shifting the extreme-mass cluster to higher values is degenerate with an increase in $\sigma_8$ (figure \ref{figothers}). The degeneracy can be broken by combining cluster data with CMB constraints.
 
\end{enumerate}

An important ingredient in our calculation is the mass function. In the presence of primordial non-Gaussianity, it remains to be seen what the correct mass function should be. Our investigation showed that the Press-Schechter, Sheth-Tormen and Tinker mass functions give similar extreme-value statistics at low redshift, but there are large differences at high redshift and large $\fnl$. The understanding of the correct form of the mass function appropriate for these extreme-mass objects is important since the uncertainty in the distribution of extreme-mass clusters due to the mass function is comparable with that from the mass determination (typically $\sigma_{\ln M}\sim0.3$). Thus, it remains for further numerical simulations along the lines of \cite{wagner,pillepich} to establish the validity of the various mass functions and non-Gaussian correction factors in the presence of non-Gaussianity.

Another crucial ingredient is the bias which, in this work, was calculated using the real-space formalism given in \cite{valageasA,valageasB} combined with our averaging procedure outlined in Appendix A. As pointed out in these papers, it is possible to extend the calculation to other types of non-Gaussianity (non-local or higher-order local type). It will be an interesting extension to study extreme-value statistics in the presence of different types of non-Gaussianity.



\mmm
\sss
\bbb

\centerline{\bb{Acknowledgment}}

\mmm

We are indebted to the anonymous referee, whose many insightful comments led to a major improvement of the paper. We are also grateful to Olaf Davis for helpful discussions in the initial stages, and to Christopher Gordon, Aseem Paranjape, Shaun Hotchkiss and, in particular, Jean-Claude Waizmann for their  comments on an early version of the manuscript. SC supported by Lincoln College, Oxford.
\bbb

\appendix 

\section{Volume-averaged bias and its approximation}

Let $b^G(M)$ be the real-space bias associated with objects of mass $M$ with $\fnl=0$. In the text, we have seen that $b^G$ is independent of $r$ to a good approximation. In the presence of non-Gaussianity, the large-$r$ behaviour of $b(M,r)$ is given by 
\be b(M,r)&\approx& f(r)g(M), \mbox{ where}\\
f(r)&\equiv& 1+K(z)\fnl\bkts{r\over h^{-1}\mbox{Mpc}}^2\lab{fr},\\g(M)&\equiv& b(M, \fnl =0),\ee
and $K(z)$ is independent of $r,$ $M$ and $\fnl$. 
This approximation allows the averaging \re{b1}-\re{b2} to be performed separately on $f(r)$ and $g(M)$.  

Our goal is to perform the averaging \re{b1} within a given volume. In the analysis of clusters lying within a redshift range $\Delta z$, the associated volume is a spherical shell whose thickness is proportional to $\Delta z$. Whilst the integration \re{b1} could, in principle, be evaluated using a 6-dimensional Monte-Carlo integration, in this Appendix we show how \re{b1} could be reduced to a triple integral. The results presented here are clearly applicable to other fields in which volume averages are required.

To begin, let us first consider an integral of the form
\be I(\mb{x_1})=\int_V \mb{dx_2}f(|\mb{x_1}-\mb{x_2}|),\lab{I}\ee
where $\mb{x_1}$ is a fixed vector and $V$ is a sphere.

\subsection{Integration within a given sphere}

If $\mb{x_1}$ lies inside a given sphere of radius $L$, we rotate the coordinate axes so that $\mb{x_1}$ lies along with the $z$-axis. We then translate the origin to the tip of $\mb{x_1}$. Let the spherical coordinates centred on this new origin be given by $(u,\theta,\phi)$. One can show that the equation of the surface of the sphere is given by
\be u = -x_1\cos\theta+\sqrt{L^2-x_1^2\sin^2\theta}, \quad x_1 = |\mb{x}_1|. \ee
Hence, the integral \re{I} can be written as
\be \int_V \mb{dx_2}\ff f(|\mb{x_1}-\mb{x_2}|) &=&\int_0^{\pi}\sin\theta d\theta  \int_{0}^{U}u^2du\int_0^{2\pi}d\phi \ff f(u,\theta,\phi),\\
&=& 2\pi\int_{-1}^{1}d\mu \int_{0}^{U}\!du \ff u^2 f(u), \lab{outer}\ee
where the integration limit is given by
\be U = -x_1\mu +\sqrt{L^2-x_1^2(1-\mu^2)}. \ee

\subsection{Integration outside a given sphere}

If $\mb{x_1}$ lies outside a given sphere of radius $\ell$, the same transformation gives

\be \int_V \mb{dx_2}\ff f(|\mb{x_1}-\mb{x_2}|) = 2\pi\int_{-1}^{-\sqrt{1-\ell^2/x^2}}d\mu \int_{u^-}^{u^+}\!du \ff u^2 f(u), \lab{inner}\ee
where 
\be u^{\pm} = -x_1\mu \pm\sqrt{\ell^2-x_1^2(1-\mu^2)}. \ee

\subsection{Averaging within a shell}

Denote $r=|\mb{x_1}-\mb{x_2}|$. The net contribution of $f(r)$ within a spherical shell of inner radius, $\ell$, and outer radius, $L$, is obtained by subtracting \re{inner} from \re{outer}. Finally, the average of $f$ over the entire shell is obtained by integrating $\mb{x_1}$ over the sphere, and then dividing by the volume of the sphere. 
\be \bar{f}\sub{shell}&=&{1\over V\sub{shell}^2}\int_V \mb{dx}_1\int_V \mb{dx}_2 \ff f(r),\lab{what}\\
 &=& {4\pi\over V\sub{shell}^2}\int_\ell^L x_1^2 dx_1 \bkts{\re{outer}-\re{inner}},\lab{shell}\ee
where
\be V\sub{shell}={4\pi\over3}(L^3-\ell^3).\ee 
 
\subsection{Applications to the bias}

Let us first apply \re{shell} to calculate the bias averaged within a sphere radius $L$ (eqs. \re{beeta}-\re{gm}). Substituting $\ell=0$ and $f(r)$ as in \re{fr} we find
\be \bar{f}\sub{sphere}=1+{6\over 5}K(z)\fnl \bkts{L\over h^{-1}\mbox{Mpc}}^2.\ee
Therefore, the bias averaged over $V$ for mass $>M$ (Eq. \ref{beta}) becomes
\be  \beta(L,M)\approx \bar{f}\sub{sphere}G(M),\ee
where
\be G(M) \equiv {1\over n(>M)}\int_M^\infty b(m, \fnl=0) {dn\over dm}\ff dm.\ee
More generally, with nonzero $\ell$, we find
\be \bar{f}\sub{shell}={L^3\bar{f}\sub{sphere}(L)+\ell^3 \over L^3 - \ell^3}.\lab{fshell}\ee
Note that by setting $\ell=0$, we recover $\bar{f}\sub{sphere}$. To include the averaging over redshift, one performs the replacement 
\be \bar{f}\sub{shell}\rightarrow{1\over V} \int_{\Delta z} dz \ff \bar{f}\sub{shell}{dV\over dz}.\ee


\section{Extremal Types}

The shape of the extreme-value distribution function holds valuable information about the statistical nature of galaxy clusters. The following theorem, which roughly states that extreme-value distributions converge to one of only three possible types, lies at the heart of extreme-value theory.

\newtheorem*{thmm}{Theorem}

\begin{thmm}[\ii{Extremal Types Theorem}] Let $X_i$ be independent and identically distributed random variables. Define the block maximum as $M_n \equiv \max_{1\leq i\leq n} \{X_i\}$. If, for some constants $a_n>0$, $b_n$, we have 
$$P(a_nM_n+b_n\leq x)\longrightarrow G(x) \quad \mbox{as }\ff n\longrightarrow \infty,$$
(in other words, if the distribution of rescaled maxima converges to a distribution $G$ for large sample size), then $G$ is one of the following distributions:\\
\ii{I. Gumbel type,} $G(x)=\exp(-e^{-y})$\\
\ii{II. Fr\'{e}chet type,} $\displaystyle G(x)=\begin{cases}0, \quad x\leq b\\ \exp(-y^{-\alpha}),\quad x > b \end{cases}$\\
\ii{III. Weibull type,} $\displaystyle G(x)=\begin{cases}\exp(-(-y)^{\alpha}),\quad x \leq b\\1, \quad x > b \end{cases}$\\
where $y=ax+b$, \ff $a,b,\alpha$ are constants, $a>0$ and $\alpha>0$. 
\end{thmm}
\sss
See, for example, \cite{leadbetter} for the proof. In this Appendix, we investigate which of these extremal types do the distributions of extreme-mass clusters belong to. 
\sss

The following function will be useful in distinguishing between the three cases:
\be g(x)=-\ln(-\ln P(x)).\ee
In the case of the Gumbel distribution, $g(x) = P^{-1}(x)=\inf\{y: P(y)\geq x\}$, which means that $g(x)$ is the \textit{$x$-quantile} of $P$. We shall refer to $g(x)$ as the quantile function \cite{jondeau,reiss}. 

To see which extremal type a given extreme-value distribution, $P(x)$, belongs to, one simply plots the quantile function and analyse its curvature for increasing patch size $L$. If the quantile is a linear, the distribution is of Gumbel type. If it concaves up (\iee $g^{\pr\pr}(x)>0$), the distribution is of Weibull type. If the quantile concaves down, it is of Fr\'{e}chet type. Note that the quantiles must be plotted on linear and not logarithmic scales.

\begin{figure}
\centering
\hskip -0.5 cm \includegraphics[width= 6.2cm, angle = -90]{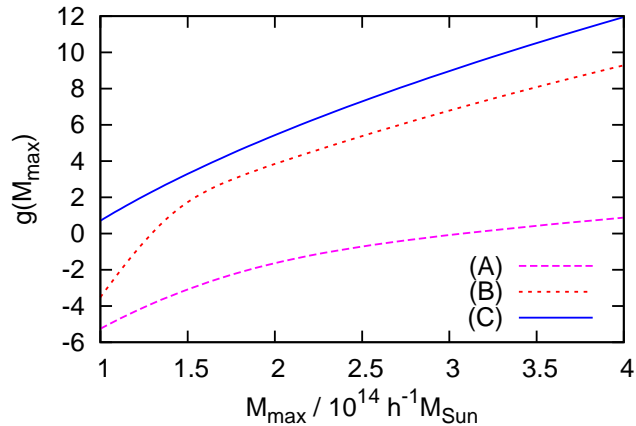}
\caption{The quantile plots for the distributions of extreme-mass clusters for the three cases: (A) $\fnl=100$, $L=100h^{-1}$Mpc, $z=1$, (B) $\fnl=200$, $L=500h^{-1}$Mpc, $z=3$, and (C) $\fnl=0$, $L=500h^{-1}$Mpc, $z=3$. The Tinker mass function was used. The concavity of these curves suggests that they belong to the Fr\'{e}chet class of distribution though they approach the Gumbel distribution at the high-mass end. This behaviour is insensitive to changes in all other parameters.}
\label{gz}
\end{figure} 


Figure \ref{gz} shows the quantile plot, $g(M\sub{max})$, of the distribution of extreme-mass clusters, $P(M\sub{max})$, with $\fnl$ in the range $0-200$. The parameters for each curve are those listed in the cases (A), (B) and (C) in section \ref{extremesubsec}, and the Tinker mass function is used. The concavity of these graphs clearly shows that the distribution of extreme-mass clusters are of the Fr\'{e}chet type, although the tails of the quantile graphs show an almost linear (\iee Gumbel) behaviour.

The Fr\'echet distribution\footnote{Some applications of the Fr\'{e}chet distribution to environmental sciences are summarised in \cite{kotz,reiss}} arises in situations when there is a natural lower limit in the distribution function ($P(x)=0$ for $x\leq$ some constant). In our case, the definition of a galaxy cluster (\eg via $M\sub{200}$) translates to a loose lower bound on $M\sub{max}$ and this may explain why the distribution of extreme-mass clusters appears to be of the Fr\'echet type. If only the high-mass tail of the distribution is taken into account, the Gumbel distribution is a reasonable approximation. As pointed out in \cite{colombi}, if the underlying distribution is exactly Gaussian, the distribution can be shown to converge to the Gumbel type, albeit very slowly. In any case, we find that $P(M\sub{max})$, for all practical purposes, belongs to the  Fr\'{e}chet family.

This conclusion is remarkably robust against changes in $\fnl$, mass function, redshift and patch size. It may be possible that this insensitivity stems from the truncation of the series \re{logp}. A more thorough approach to studying the extremal-type convergence is to fit the distribution to some functional form (\eg see \cite{mikelsons,harrison2} in which the extreme-value distributions are modelled as a generalised extreme-value distribution) or prove the convergence using one of the criteria given, for example, in \cite{kotz, leadbetter}. Like \cite{mikelsons,harrison2}, we find the convergence insensitive to the value of $\fnl$.

We note that, contrary to the observation in \cite{colombi}, we found no combination of parameters which give rise to a Weibull distribution, which arises when there is a natural upper bound for the distribution function. Moreover, it is worth noting that if the pdfs such as those in figures \ref{figbig} and \ref{varyL} are well-approximated by `skew-symmetric' functions (\eg an Edgeworth expansion) then the distribution cannot converge to the Weibull type as proven in  \cite{chang}.

Finally, we point out an interesting fact that if the coefficients in the expansion \re{white} conspire to make $P(x)$ an exactly Poissonian distribution
\be P(x,\lambda)=e^{-\lambda}\sum_{k=0}^x {\lambda^k\over k!},\ee
then the limiting distribution $G(x)$ will completely degenerate to $G=1$ or $0$. This is one of the rare examples where the extreme-value distribution does not converge to any of the three standard distributions. Of course, we do not expect a realistic distribution of galaxy clusters to be exactly Poissonian.

\bibliographystyle{h-physrev3}
\bibliography{gumbelrevised}

\end{document}